\documentclass[prx,superscriptaddress,longbibliography,twocolumn]{revtex4-1}
\usepackage{graphicx}
\usepackage{bm}
\usepackage{xfrac}
\usepackage{xcolor}
\usepackage{physics}
\usepackage{array}
\usepackage{amsmath}
\usepackage{appendix}
\usepackage{float}

\newcolumntype{P}[1]{>{\centering\arraybackslash}p{#1}}

\newcommand{\be}{\begin{equation}}
\newcommand{\ee}{\end{equation}}

\newcommand{\bQ}{{{\bf{Q}}}}

\newcommand{\bea}{\begin{eqnarray}}
\newcommand{\eea}{\end{eqnarray}}
\newcommand{\beal}{\begin{align}}
\newcommand{\eeal}{\end{align}}

\newcommand{\upa}{\uparrow}
\newcommand{\dna}{\downarrow}

\newcommand{\dg}{{\dagger}}
\newcommand{\pdg}{{\phantom\dagger}}

\begin{document}
%

\title{$XY$ magnetism, Kitaev exchange, and long-range frustration in the $J_{\rm eff}=1/2$ honeycomb cobaltates}

\author{Shreya Das}
\thanks{These authors contributed equally}
\affiliation{Department of Condensed Matter Physics and Materials Science, S.N. Bose National Centre for Basic Sciences, Kolkata 700098, India.}

\author{Sreekar Voleti}
\thanks{These authors contributed equally}
\affiliation{Department of Physics, University of Toronto, Toronto, Ontario, M5S 1A7, Canada}

\author{Tanusri Saha-Dasgupta}
\thanks{tanusri@bose.res.in}
\affiliation{Department of Condensed Matter Physics and Materials Science, S.N. Bose National Centre for Basic Sciences, Kolkata 700098, India.}

\author{Arun Paramekanti}
\thanks{arunp@physics.utoronto.ca}
\affiliation{Department of Physics, University of Toronto, Toronto, Ontario, M5S 1A7, Canada}

\begin{abstract}
The quest for
Kitaev quantum spin liquids has led to great
interest in honeycomb quantum magnets with strong spin-orbit coupling.
It has been recently proposed that even Mott insulators with $3d$ transition metal ions, having nominally weak spin-orbit coupling, can realize such exotic
physics. Motivated
by this, we study the rhombohedral
honeycomb cobaltates CoTiO$_3$, BaCo$_2$(PO$_4$)$_2$, and BaCo$_2$(AsO$_4$)$_2$, using
{\it ab initio} density functional theory, which takes into account realistic crystal field distortions and chemical information,
in conjunction with exact diagonalization numerics.
We show that these Co$^{2+}$  magnets host $J_{\rm eff}\!=\! 1/2$
local moments with highly
anisotropic $g$-factors, and we extract their
full spin Hamiltonians including longer-range and anisotropic exchange couplings. For
CoTiO$_3$, we find a nearest-neighbor easy-plane ferromagnetic $XXZ$ model
with additional bond-dependent anisotropies and interlayer exchange,
which supports three-dimensional (3D) Dirac nodal line
magnons.
In contrast, for BaCo$_2$(PO$_4$)$_2$ and BaCo$_2$(AsO$_4$)$_2$, we find a strongly suppressed interlayer coupling, and
significant frustration from additional third-neighbor antiferromagnetic exchange mediated by P/As. Such bond-anisotropic $J_1$-$J_3$
spin models can support collinear zig-zag or coplanar spiral ground states. We discuss their dynamical spin correlations which reveal a gapped
Goldstone mode, and argue that the effective
parameters of the pseudospin-$1/2$ models in these two materials may be strongly renormalized by coupling
to a low energy spin-exciton. Our results call for re-examining proposals for realizing Kitaev spin liquids
in the honeycomb cobaltates.
\end{abstract}

\maketitle

\section{Introduction}
A significant effort has been invested in exploring material realizations of Kitaev's honeycomb quantum spin liquid (QSL) model and its anyon
excitations \cite{Kitaev2006,Kimchi2018,Takagi2019,Motome2020,takayama2021}. The initial work in this direction focussed
mainly on the iridate honeycomb magnets which have strong spin-orbit coupling (SOC) \cite{Chaloupka2010,Okamoto2007,
Katukuri2014,Rau2014,Singh2010,
Choi2012,Kimchi2011}. At this point, the
most promising candidate appears to be $\alpha$-RuCl$_3$ \cite{Plumb2014,Sears2015,Banerjee2016,Weber2016,Gronke2018,Mashhadi2018,Tian2019,Zhou2019,Mashhadi2019,Biswas2019,Gerber2020},
where intermediate in-plane Zeeman fields
appear to lead to a plateau in the thermal Hall conductivity
$\kappa_{xy}$ \cite{Kasahara2018},
and quantum oscillations in the diagonal thermal conductivity $\kappa_{xx}$ \cite{Czajka2021},
which have been proposed as potential signatures of emergent Majorana fermion
excitations in an insulating quantum magnet.
At the same time, there is debate on the precise magnetic Hamiltonian
for $\alpha$-RuCl$_3$ \cite{Chernyshev2020,suzuki2020} as well as the existence of a field induced QSL
\cite{GegenwartPRL2020,GegenwartPRB2021} in its phase diagram.
A search for such exotic physics in a wider range of quantum materials is thus highly desirable.

This quest has led to great interest in honeycomb cobaltates, where the Co$^{2+}$
ion in an octahedral crystal field environment has
total spin $S=3/2$ moment
and an effective total orbital angular momentum $L=1$, which are locked by spin-orbit coupling, leading to a
$J_{\rm eff}\!=\!1/2$ pseudospin doublet
ground state \cite{Liu2018,Sano2018,Liu2020}.
These materials were proposed to host to dominant Kitaev exchange between neighboring pseudospins,
rendering them candidates for quantum spin liquids \cite{Liu2018,Sano2018,Liu2020,Liu2021}.
While this is an exciting proposal, SOC is much weaker for $3d$ transition metal ions as compared with the
$5d$-iridates or $4d$-ruthenates. As a result,
trigonal distortions of the local crystal field
environment, inevitable in any layered honeycomb material,
may be expected to have a considerable impact on the nature of exchange interactions and the low energy
fate of magnetism in these materials.

\begin{figure}
  \centering
  \includegraphics[width=.46\textwidth]{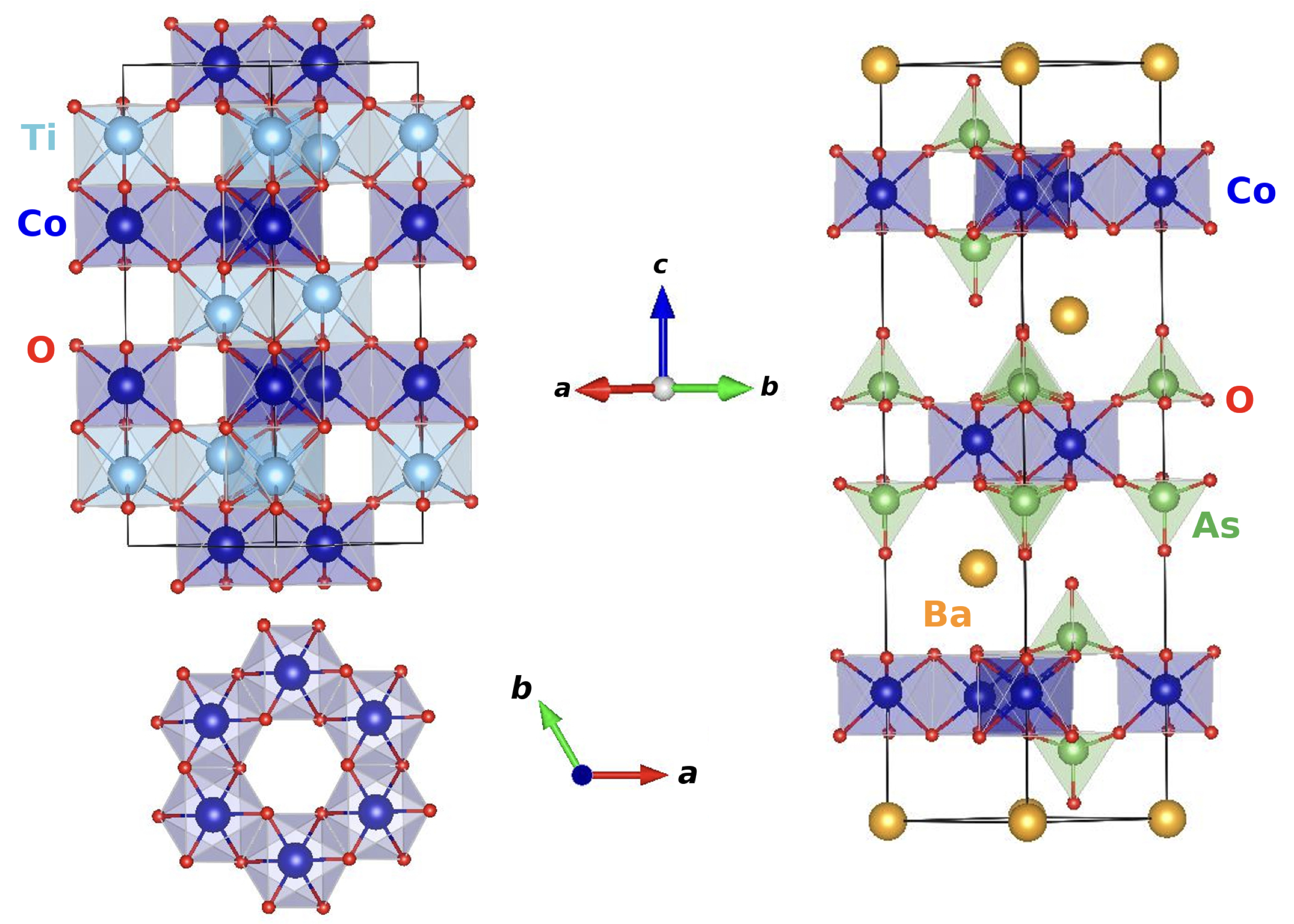}
  \caption{Crystal structure of CoTiO$_3$ (left) and BaCo$_2$(AsO$_4$)$_2$ (right) viewed with hexagonal c-axis pointing
    upwards. The Co, Ti, As, Ba and O atoms are labeled and colored differently. Marked is the
    six formula unit hexagonal unit cell. Bottom left panel shows the top view of CoTiO$_3$ looking down the $c$-axis.}
  \label{fig:structure}
\end{figure}

To critically examine this issue, we explore several rhombohedral $d^7$ cobaltates --- CoTiO$_3$, BaCo$_2$(PO$_4$)$_2$, and
BaCo$_2$(AsO$_4$)$_2$ --- which are formed from stacked
honeycomb Co$^{2+}$ layers as shown in Fig.~\ref{fig:structure}. In particular, we ask: does a realistic
theory of the $d^7$ cobaltates, using a combination of {\it ab initio} density functional theory (DFT) and exact diagonalization numerics, favor the
realization of Kitaev spin liquids in these materials?

To address this issue, we use non-spin-polarized DFT calculation to extract the non-interacting part of the Hamiltonian, namely the hopping interactions and crystal field splitting, and then solve the many-body problem of the Co multiplet structure by including Coulomb interactions and spin-orbit coupling in an exact diagonalization for a single Co ion. Finally we consider the lowest two $d^7$ states as the pseudospin-1/2 basis, to derive superexchange interactions by utilizing the multiplet structure and incorporating hoppings within second-order perturbation theory. Our approach thus combines the material specific theory DFT with the many-body
calculation to handle the multiplet structure, which we use to shed light on magnetism in the $d^7$ cobaltates.

A quick summary of our key results are as follows.
Our {\it ab initio} computations reveal that although all three compounds host the stacked
honeycomb Co$^{2+}$ layers, the difference in the geometry and chemical composition of the spacer layers between the Co$^{2+}$ layers greatly
influences the resulting electronic structure of the three compounds. In CoTiO$_3$, strong Co-Ti covalency leads to a more 3D
electronic structure, while the delicate Co-As/P covalency and large separation between the Co$^{2+}$ layers leads to weakly coupled 2D layers
but with longer-range hoppings in the electronic structure of BaCo$_2$(AsO$_4$)$_2$ and BaCo$_2$(PO$_4$)$_2$. We use
these DFT
inputs to carry out an exact diagonalization study incorporating correlation effects, which has been shown to work well in other
Mott insulators with SOC \cite{Winter2016,Biswas2019}.
Our exact diagonalization numerics reveal that the Co$^{2+}$ ion has strongly anisotropic $g$-factors, consistent with experimental data
\cite{Zhong2020,Yuan2020,Elliot2020}.
For
CoTiO$_3$, our computations reveal a dominant nearest-neighbor $XXZ$ easy-plane ferromagnetic coupling, with additional anisotropic compass-type
exchange interactions and interlayer couplings. Such a spin Hamiltonian supports 3D Dirac nodal line magnon
excitations and is in
good quantitative agreement with recent neutron scattering experiments \cite{Yuan2020,Elliot2020}. For BaCo$_2$(PO$_4$)$_2$ and
BaCo$_2$(AsO$_4$)$_2$, we find that the spacer layers lead to nearly decoupled honeycomb layers. The magnetic Hamiltonian is found to be a
quasi-2D easy-plane $XXZ$ model with significant frustration from third-neighbor antiferromagnetic exchange. Such $J_1$-$J_3$
honeycomb lattice spin models
can support collinear zig-zag or coplanar spiral ground states \cite{RASTELLI1979,J.B.FouetP.SindzingreC.Lhuillier2004,Kimchi2011}.
We present results on their dynamical spin structure factor
which reveals a gapped Goldstone mode from bond-dependent compass anisotropy terms, in qualitative agreement with neutron
scattering data \cite{Nair2018,Regnault2018}.

\section{Crystal structure} \label{sec:crystalstructure}

All three compounds CoTiO$_3$, BaCo$_2$(AsO$_4$)$_2$, and BaCo$_2$(PO$_4$)$_2$ crystallize in the rhombohedral $R\bar{3}$ space group.
While the basic structural motif of all three compounds consist of stacked honeycomb cobaltate layers formed by edge shared
CoO$_6$ octahedra, the out of plane stacking involving spacer layers between honeycomb planes are distinctly
different between CoTiO$_3$, and BaCo$_2$(AsO$_4$)$_2$ or BaCo$_2$(PO$_4$)$_2$. The left and middle panels of Fig. 1
show the crystal structures of CoTiO$_3$ and BaCo$_2$(AsO$_4$)$_2$, respectively; the crystal structure of
BaCo$_2$(PO$_4$)$_2$ being very similar to that of BaCo$_2$(AsO$_4$)$_2$, is not shown. In case of CoTiO$_3$, the cobalt
layers are separated by Ti layers, where octahedrally coordinated Ti$^{4+}$ ions face share with CoO6$_6$ octahedra along the
hexagonal $c$-axis, with the Ti ion on top of Co ion, vertically displaced by $2.67${\AA}. In case of BaCo$_2$(AsO$_4$)$_2$,
on the other hand, the Co layers are separated by bilayers of
AsO$_4$ tetrahedra, one pointing up
and another pointing down, and separated by large Ba$^{2+}$ ions. This difference in spacer layer thickness between
CoTiO$_3$ and the Ba compounds, makes the $c$-axis lattice constant of BaCo$_2$(AsO$_4$)$_2$ and BaCo$_2$(AsO$_4$)$_2$
significantly different from CoTiO$_3$, being $13.92${\AA} for  CoTiO$_3$, $23.49${\AA} for BaCo$_2$(AsO$_4$)$_2$, and  $23.22${\AA}
for BaCo$_2$(PO$_4$)$_2$. However, their
inplane lattice constants are rather similar: $5.06${\AA} for  CoTiO$_3$, $5.00${\AA} for BaCo$_2$(AsO$_4$)$_2$, and $4.86${\AA} for
BaCo$_2$(PO$_4$)$_2$. This gives rise to a more 2D character of the Co network in Ba compounds, while CoTiO$_3$ has more 3D character.
Furthermore, as opposed to face-shared geometry of Co-Ti network, the AsO$_4$ tetrahedra corner share with
CoO$_6$ octahedra, the As ion being displaced from Co ion by $1.67${\AA} along the in-plane directions and $1.41${\AA} along the vertical direction.
This creates O-As-O bridges both from top and bottom between long ranged in-plane Co neighbors, this feature being absent for CoTiO$_3$ compound.
These structural details have important implications for the electronic structure and low-energy Hamiltonian
of the three compounds which we discuss next.

\section{Electronic Structure}

To characterize the valences, spin states, and orbital contributions of various ions, we first carry out the spin-polarized
DFT calculations within the framework of generalized gradient approximation (GGA) \cite{GGA}.
See Supplementary Materials (SM) \cite{SuppMat} for calculation details \cite{vasp,PAW,PBE}.
The magnetic moment
of Co site is found to be around $2.60 \mu_B$ confirming the high spin $d^7$ configuration of Co ions. Interestingly for CoTiO$_3$,
large fraction of the missing moment is found to be at Ti and O sites of $0.12 \mu_B$ and $0.10 \mu_B$, respectively, despite the
nominally nonmagnetic configuration of Ti$^{4+}$ and O$^{2-}$, indicating strong
Co-O, and Ti-O covalency. For the Ba compounds, a similar magnitude of missing moment ($\sim\! 0.10 \mu_B$) is found at O sites.
Furthermore, a non-negligible moment of $\sim\!0.05\mu_B$ is also found at the nominally valenced As$^{5+}$/P$^{5+}$ sites,
 indicating the subtle role of As/P in describing the electronic structure of the Ba compounds. Repeating the calculations
within the GGA+SOC approach uncovers a substantial orbital moment
 of $0.16$-$0.17 \mu_B$ at the Co site highlighting importance of SOC in these compounds.

 In order to check
 the effect of missing correlation effect beyond GGA, GGA with supplemented Hubbard $U$ correction \cite{GGAUSOC} within the framework
 of GGA+$U$ and GGA+SOC+$U$ was carried out with choice of $U\!=\! 3$\, eV and $J_H \!=\! 0.7$\,eV \cite{WangU2006,Cortada2016}. We find that the
 resulting magnetic moments remain qualitatively unchanged; however, inclusion of a supplemented Hubbard $U$ correction strongly impacts the electronic structure.
Crucially, an insulating solution is only obtained upon addition of $U$ correction, while GGA and GGA+SOC solutions are metallic. The magnetic moments
   calculated within GGA, GGA+SOC and GGA+SOC+$U$ as well as density of states calculated within three scheme of
   calculations are given in the SM \cite{SuppMat}.

\begin{figure}[t]
  \centering
  \includegraphics[width=.46\textwidth]{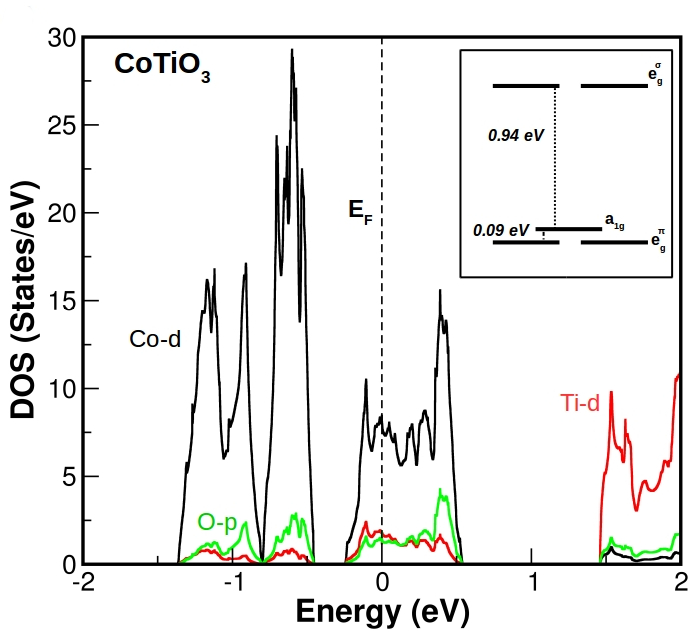}
  \includegraphics[width=.46\textwidth]{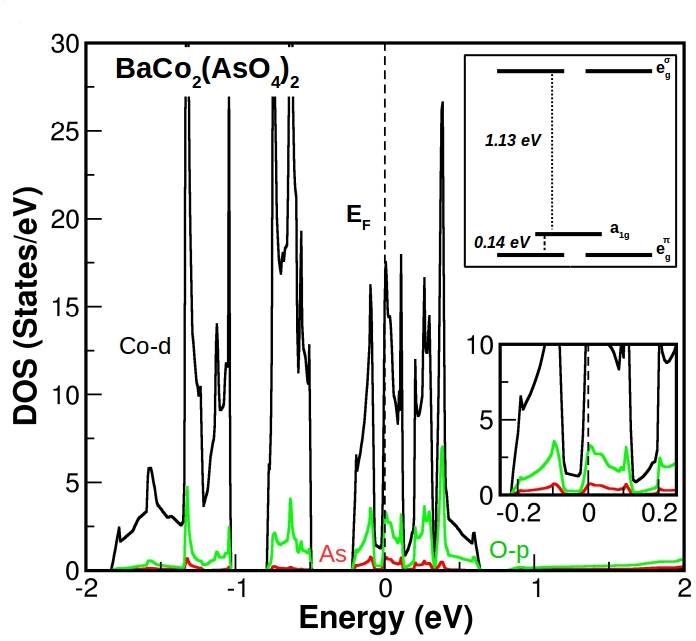}
  \caption{Non spin-polarized DFT density of states of CoTiO$_3$ (top) and BaCo$_2$(AsO$_4$)$_2$ (bottom) projected
    to Co-$d$, Ti-$d$, As and O-$p$ states. The zero of the energy is marked at the corresponding Fermi level (E$_F$).
    Crystal field splitting of the Co-$d$ states are shown as insets. The second inset in bottom is the
  zoomed-in plot of the density of states close to E$_F$.}
  \label{fig:DOS}
\end{figure}

 Since our goal is to use the DFT calculations to derive a low-energy Hamiltonian to be used as an input for a model Hamiltonian study,
 we henceforth focus on the non-spin-polarized electronic structure of these compounds.
 Fig.~2 shows the density of states of CoTiO$_3$ and BaCo$_2$(AsO$_4$)$_2$ compounds
 projected to different orbital characters; the electronic structure of BaCo$_2$(PO$_4$)$_2$ is very
 similar to  BaCo$_2$(PO$_4$)$_2$ and is thus not shown. In an ideal octahedral environment the Co $d$ states are split into three-fold
 degenerate $t_{2g}$ and two-fold degenerate $e_g$ states. However, the CoO$_6$ octahedra in the studied compounds are not perfect,
 rather they are trigonally distorted with the Co atom moved along [111] direction from the centre of the octahedra, leading to
 three short and three long Co-O bonds. As shown in the insets of Fig.~2, this further splits the $d$ levels into two-fold degenerate $e_g^\pi$,
 non-degenerate $a_{1g}$, and two-fold degenerate $e_g^\sigma$ levels. The octahedral splitting is found to be about $\sim\! 1$ eV, while trigonal
 splitting is found to be $\sim\! 0.1$\,eV and thus larger than the nominal SOC on Co. Within the non-spin-polarized scheme
 of calculations, the Co $t_{2g}$ are fully occupied while the Co $e_g$ states are partially filled. As seen from the density of states
 near the Fermi level, there is significantly contribution
 of O $p$ states along with the Co $d$ states, indicating strong Co-O covalency; this was also manifest in the presence of substantial
 magnetic moment at the otherwise non-magnetic O sites. In CoTiO$_3$, we also notice a significant presence of Ti-$d$ character
 near the Fermi level, which we deem responsible for the 3D nature of the density of states. While the density of states of
 BaCo$_2$(AsO$_4$)$_2$ bears an overall resemblance to that of CoTiO$_3$, the details are markedly different. Near the
 Fermi level, we find the appearance of pseudogap like features seen in the inset of the bottom panel of Fig.~2, which arises
 due to intricate interplay of long-ranged in plane Co-Co interactions mediated by As; this feature is absent in CoTiO$_3$.
 While the contribution of As states near the Fermi level is small compared to Ti, it is still non-negligible as highlighted in the
 zoomed-in plot of the density of states in the second inset of the bottom panel of Fig.~2. As discussed below, we believe this overlap with As states
 drives the overlap of further neighbor Wannier orbitals in the Ba compounds and leads to strong third-neighbor exchange.

 \section{Low-energy DFT Hamiltonian}

 \begin{figure}
  \centering
  \includegraphics[width=.46\textwidth]{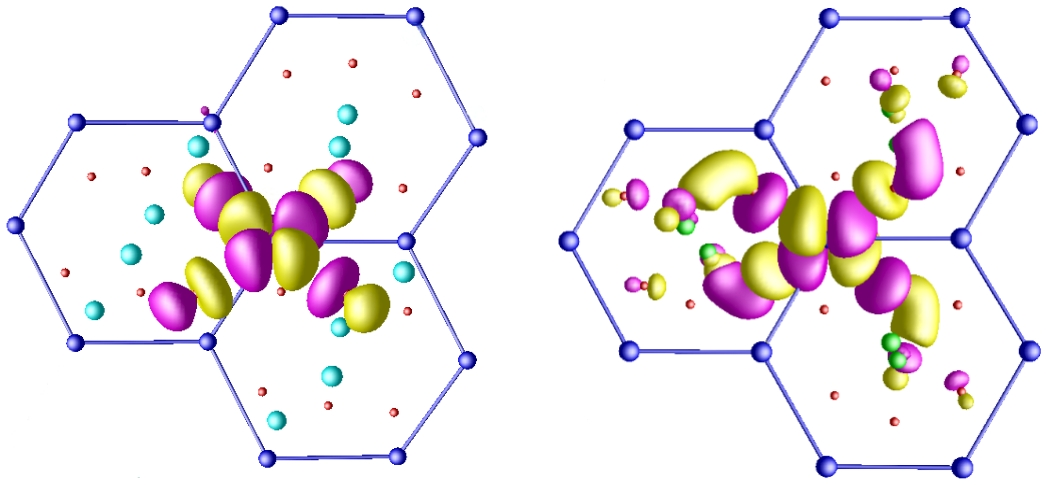}
  \caption{In-plane view of NMTO downfolded Co $d_{x^2-y^2}$ Wannier function of CoTiO$_3$ (left) and BaCo$_2$(AsO$_4$)$_2$ (right).
    Plotted are the constant value isosurfaces with lobes of opposite signs colored as magenta and yellow. The hexagonal network of
   Co atoms in blue is shown, with O and Ti/As atoms shown as small-red and medium-cyan/green balls.}
  \label{fig:wannier}
\end{figure}

 In an attempt to derive Co $d$-only low energy Hamiltonian out of DFT calculations, we resort to the energy selective
 N-th order muffin-tin orbital (NMTO) based downfolding technique \cite{NMTO}. Starting from a self-consistent DFT calculation
 in linear muffin-tin orbital
 basis,\cite{LMTO}, NMTO-downfolding
 calculation arrives at a low-energy Co $d$-only Hamiltonian by integrating out degrees that are not of interest like O-$p$, Ti-$d$,
 As and Ba states. This procedure constructs the ab-initio derived effective Co Wannier functions, the head parts of which are shaped according
 to Co $d$ symmetries, while the the tail parts are shaped according to the integrated out degrees of freedom which are
 mixed in with Co $d$ characters. The construction of these effective Wannier functions depends on the choice of basis.
 We chose the octahedral local coordinate system with $z$-axis of the local coordinate
 system pointed along one of short Co-O bond, and the local y-axis pointed closet to neighboring Co-O bond. With this
 choice, the local $z$-axis on nearest neighbor edge-shared Co pairs point in opposite directions, with the center of inversion
at the middle of the vector connecting the Co-Co pair. Fig.~\ref{fig:wannier} shows a representative example of the effective Co $d_{x^2-y^2}$  Wannier
function as given by NMTO-downfolded calculations for CoTiO$_3$ and BaCo$_2$(AsO$_4$)$_2$, which exhibits $pd\sigma$
 anti-bonds formed between Co and O, which appear as pronounced tails shaped according to O $p$ symmetry. In the case of CoTiO$_3$,
 a significant weight of the tail also sits at the out-of-plane Ti site, as we find from other Wannier orbitals (not shown). What is remarkable
 in case of Ba compound is that one of the lobes of the O-tail bends towards As sites due to overlap of Co-O and As-O covalency
 at the corner-shared O site of CoO$_6$ octahedra and AsO$_4$ tetrahedra. This As-covalency driven bending of the tails
 promotes the long-ranged third neighbor in-plane Co-Co interaction, which plays an important role in the spin
 model for BaCo$_2$(AsO$_4$)$_2$; we find similar features in BaCo$_2$(PO$_4$)$_2$.

 The low-energy tight-binding Hamiltonian, in the basis of NMTO-downfolded effective Co $d$ Wannier functions, encodes
 the effect of local distortion of CoO$_6$ octahedra, as well as the effect of spacer chemistry and geometry. The full
 local crystal field matrix and the hopping matrices for various neighbors are given in the SM \cite{SuppMat}.
 This information is next used to build model Hamiltonians for these cobaltates.

\section{Exact diagonalization results: Single ion properties}
From our DFT calculations, we extract a local crystal field Hamiltonian $H_{\rm CF}$ describing the d-orbital manifold of the
Co$^{2+}$ ion in a distorted octahedral crystal field environment which only has a residual $C_3$ symmetry. We supplement this
Hamiltonian with SOC and interactions, so that the on-site Hamiltonian takes the form
\bea
H_{\rm loc} = \xi H_{\rm CF} + H_{\rm SOC} + H_{\rm int}.
\label{ham}
\eea
Here, $H_{\rm SOC}$ is the single-particle SOC term, explicitly
given by
\bea
H_{\rm SOC}= \lambda \sum_{\ell \ell' m} \sum_{\alpha\beta}
c^\dg_{\ell s} L^m_{\ell \ell'}
c^\pdg_{\ell' s'} \sigma^m_{s s'}~,
\eea
where $(\ell, \ell')$ are orbital labels corresponding to
$\{ (yz,xz,xy),(x^2\!-\!y^2 , 3z^2\!-\!r^2) \}$, $m$ labels the
vector component of the orbital $L$ or spin $\sigma$ angular momentum,
with $s,s'$ being
spin component labels.
$H_{\rm int}$ is the spherically symmetric
Kanamori Hamiltonian encoding interactions
between the electrons:
\bea
\label{interactionham}
H_{\rm int} &=&
U \sum_\ell n_{\ell\upa}n_{\ell\dna} + U'\sum_{\ell > \ell'} n_{\ell}n_{\ell'} \nonumber \\
   &-& J_H \sum_{\ell\neq\ell'}
   \mathbf{S}_\ell \cdot \mathbf{S}_{\ell'} +
   J_H \sum_{\ell\neq\ell'} c^\dagger_{\ell \upa}
   c^\dagger_{\ell \dna} c^\pdg_{\ell' \dna}c^\pdg_{\ell' \upa}~.
\eea
Here, $U$, $U'$, and $J_H$ represent intra-orbital Hubbard,
inter-orbital Hubbard, and Hund's couplings, respectively, and the spin operator
in orbital $\ell$ is
\be
\mathbf{S}^m_\ell = \frac{1}{2} c^\dagger_{\ell s} \sigma^m_{s s'}
c^\pdg_{\ell s'}.
\ee
We assume $U'\!=U\!-\! 2J_H$ as appropriate for a spherically symmetric Coulomb interaction
\cite{Georges2013}.
Finally, in this single-ion Hamiltonian Eq.~\ref{ham}, we have
incorporated a scaling factor $\xi$
multiplying the $H_{CF}$ obtained from our DFT calculations,
as a crude way to account for the
``double-counting'' of interactions. We refer here to the fact that
\textit{ab initio} calculations partially
account for inter-orbital repulsion and changes in the
corresponding orbital occupancies and energies at mean
field level, while such a renormalization of crystal field
levels will also occur from the full set of Kanamori interactions
in Eq.~\ref{interactionham}.
An intrinsic issue of any realistic many-body approach is the so-called double-counting of
  interaction terms. How to choose the double-counting potential in a manner that is both physically sound and
  consistent is unknown. Various different schemes of double-counting have been proposed in literature, as it is
  an ill-posed question to make connection between DFT based on electronic densities, and a Hubbard-like
  many-body model based on localized atomic orbitals. The double counting schemes like around
  mean-field \cite{czyzyk94} or fully localized limit, \cite{Anisimov97} or self-interaction correction \cite{Seo2007} used in DFT+$U$ approach
  may provide different results. An alternative approach is to use double-counting correction as an adjustable parameter, $\xi$,
  which we have used, to give the best comparison with experiment (see discussions later).

Below, we set
$\lambda \!=\! 80$\,meV for Co$^{2+}$ \cite{Maekawa2004},
and choose the scaling factor $\xi \!=\! 0.65$.
Based on an exploration of how
the different exchange couplings vary with interactions
(discussed below),
we fix the interaction strengths to $U=3.25$\,eV
and $J_H=0.7$\,eV \cite{LJdeJongh2013}. We
find that this set of parameters leads to reasonable results
for the single ion g-factors, as well as the low energy
single ion excitations and spin-wave modes
observed in CoTiO$_3$ using inelastic
neutron scattering experiments \cite{Yuan2020,Elliot2020}.
We have found that the exchange Hamiltonians we derive in
later sections are not substantially impacted by
the precise value of $\xi$.

\subsection{Kramers doublet ground state}

As discussed in Ref~\cite{Liu2018,Sano2018}, the $3d^7$ Co$^{2+}$ ion in an
octahedral crystal field environment is predominantly in the high-spin $S\!=\!3/2$ ($t_{2g}^5 e_g^2$) configuration. Combining this with the total effective
orbital angular momentum $L\!=\!1$, of a single hole
in the $t_{2g}$ orbitals,
the SOC term with $\lambda>0$ stabilizes a $J_{\rm eff}\!=\!1/2$ pseudospin
degree of freedom. Indeed, our exact diagonalization of the
Hamiltonian leads to a Kramers doublet ground state for a wide range of parameter
values, indicating that such a pseudospin-$1/2$
picture remains valid for all three cobaltates
even in the presence of realistic distortions.

\subsection{Single-ion excitations}

Inelastic neutron scattering experiments on CoTiO$_3$
reveal magnetic excitations above the $J_{\rm eff}=1/2$ doublet at energies
$\{29(2),58(7),\ldots,132(3)\}$\,meV, where ``$\ldots$''
indicates a broad continuum of significant scattering intensity in the
$60$-$120$\,meV range where individual sharp peaks cannot be
resolved \cite{Yuan2020,Elliot2020}.
Choosing the scaling factor $\xi\!=\!0.65$ in $H_{\rm loc}$, we
find that these low-lying many-body crystal field
excitations above the
$J_{\rm eff}\!=\! 1/2$ doublet ground state in CoTiO$_3$ are at energies
$\{ \underline{22} , \underline{63} , 104 , 111 , \underline{132} \}$\,meV, with the underlined excitations energy levels being in
reasonably good agreement with the experimentally observed peaks.
We contrast these results with the corresponding
values when the scale factor is set to
$\xi\!=\! 1$, which leads to excitation energies
$\{ 15 , 84 , 118 , 130 , 152\}$\,meV quite unlike the data.

Using this scaling factor $\xi \!=\! 0.65$
for BaCo$_2$(AsO$_4$)$_2$ and BaCo$_2$(PO$_4$)$_2$, we find the excited states at excitation energies which are respectively
at $\{ 14 , 85 , 120 , 134 , 158 \}$ and $\{ 16 , 77 , 112, 126 , 151 \}$. These could
be explored in future experiments. We note that similar spin exciton states have also been measured in certain
pyrochlore cobaltates \cite{ross2017}.

\subsection{g-factors}

Table \ref{table:gfactortable} lists the $g$-factors associated
with this pseudospin degree of freedom which we
obtain from the ground state expectation value
$2 \langle (\vec L + 2 \vec S) \rangle$
in the presence of a weak Zeeman field $\vec B$.
By choosing $\vec B$ to point perpendicular or parallel to
the $C_3$ axis (i.e., $c$-axis),
we extract $g_\perp$ and $g_\parallel$ respectively.
In contrast to
the isotropic $g$-factors obtained for the iridate $d^5$ Mott
insulators, all the cobaltates studied here show highly
anisotropic $g$-factors, in agreement with recent experiments
\cite{Zhong2020,Regnault2018}. Such anisotropic $g$-factors have also been
measured in certain pyrochlore cobaltates \cite{ross2017}.
To the best of our knowledge,
$g$-factors for BaCo$_2$(PO$_4$)$_2$ have not been measured;
our estimates may be compared with
future experiments.

\begin{table}[t]
\centering
\begin{tabular}{|P{3cm}|P{2cm}|P{2cm}|}
	\hline
	\textbf{Material} & $g_{\perp}$ & $g_\parallel$ \\
	\hline\hline
	CoTiO$_3$ & 5.0 & 3.2 \\
	\hline
	BaCo$_2$(AsO$_4$)$_2$ & 5.0 & 2.7 \\
	\hline
	BaCo$_2$(PO$_4$)$_2$ & 5.0 & 3.0 \\
	\hline
\end{tabular}
\caption{Computed $g$-factors corresponding to directions perpendicular (in-plane) or parallel (out-of-plane)
to the $C_3$ axis ($c$-axis) for honeycomb cobaltates studied in this paper.}
\label{table:gfactortable} 
\end{table}

\section{Perturbation theory for inter-site exchange}

In order to obtain the exchange interactions between pseudospins
on a pair of sites,
we use the single-ion energy levels and eigenfunctions and
incorporate inter-site hopping to second order in perturbation
theory. We outline this below, together with a brief discussion
about the dependence on the various exchange couplings on interaction parameters such as $U$ and $J_H$, with further
details given in the SM \cite{SuppMat}.

\subsection{Two-site perturbation theory}

\begin{figure}
  \centering
  \includegraphics[width=.48\textwidth]{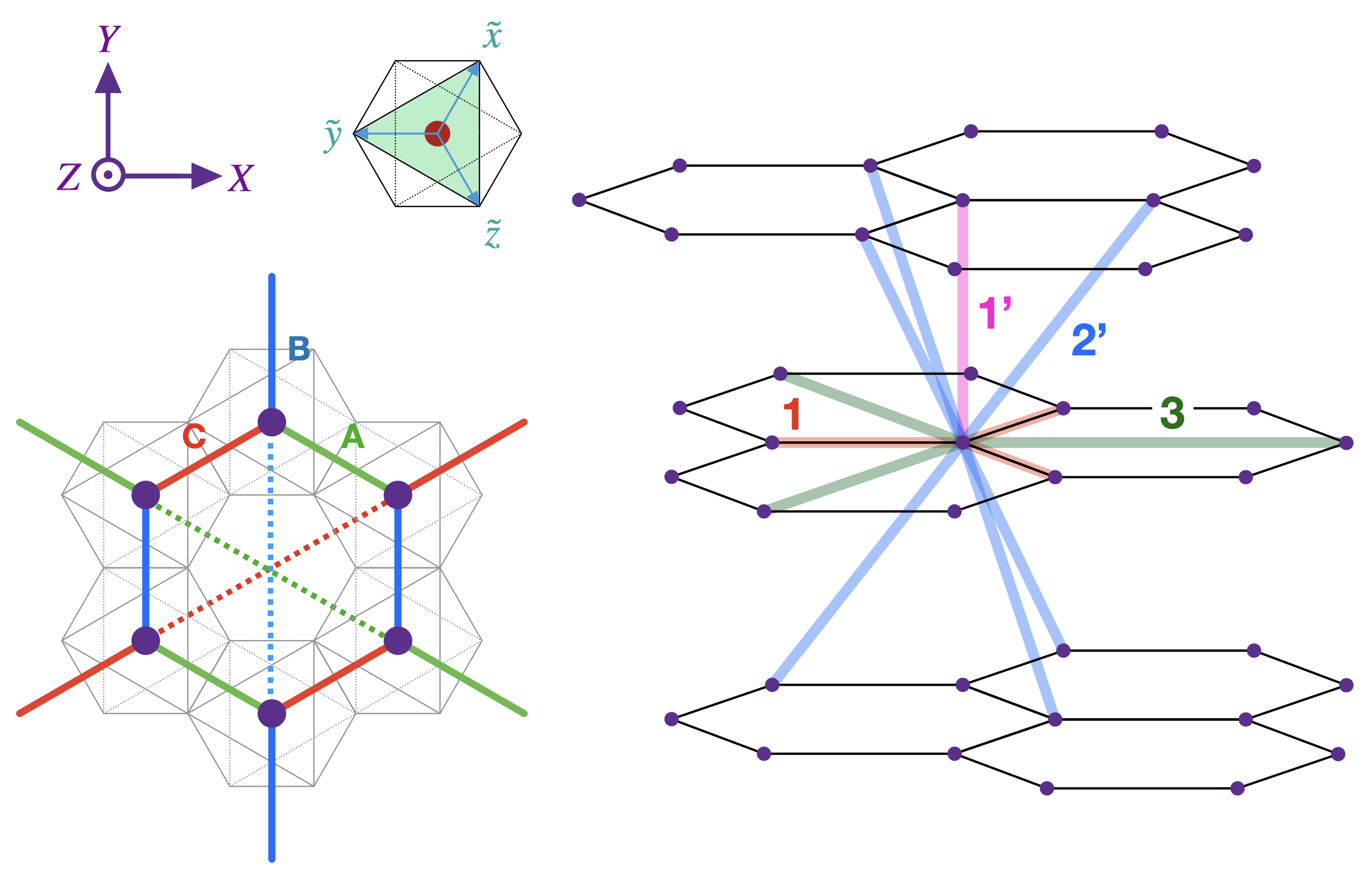}
  \caption{(Left) The honeycomb plane is shown, as viewed down the $c$ axis, along with the global $XYZ$ and local $\tilde{x}\tilde{y}\tilde{z}$ basis. The definition of the bases, along with that of the $A$, $B$, and $C$ bonds is consistent with Ref.~\cite{Chaloupka2010}. The dotted bonds denote the third-nearest neighbor bonds, which can be also be defined as $A$, $B$, $C$-type.
  (Right) Bonds with non-negligible exchange couplings for the different materials; unprimed numbers refer to in-plane neighbors, while primed numbers refer to the out-of-plane neighbors.}
  \label{fig:spinbases}
\end{figure}

The interactions between the pseudospins can be obtained via second order perturbation theory, using hopping integrals extracted
from the above DFT calculations. The calculation schematically proceeds as follows. (i) We solve the single site problem for
occupancies $d^6$, $d^7$ and $d^8$, to obtain
the full set of energy levels and many-body wavefunctions
including SOC. In doing this, we apply a weak field along the
global $Z$-axis ($C_3$ axis perpendicular to the honeycomb
planes) in order to define our pseudospin $\ket{\uparrow},\ket{\downarrow}$ basis states (see Fig.~ \ref{fig:spinbases}, reproduced from \cite{Chaloupka2010}).
(ii) We extract spin-independent hopping
matrices $T^{(ij)}$ from DFT, whose elements
$T^{(ij)}_{\ell\ell'}$ describe an electron in orbital
$\ell$ at site $i$ hopping to an orbital $\ell'$ at site $j$,
so that the $(ij)$ kinetic energy is
\bea
\hat{T}_{ij} = \sum_{\ell\ell'\alpha}(T^{(ij)}_{\ell\ell'} c^\dg_{j\ell'\alpha} c^\pdg_{i\ell\alpha} + T^{(ji)}_{\ell'\ell} c^\dg_{i\ell\alpha} c^\pdg_{j\ell'\alpha})
\eea
(iii) We focus on initial and final states where
each of the two sites live in a ground doublet which we
denote as $\ket{b}, \ket{b'}$. These kets span a $4$-dimensional
Hilbert space, formed by the direct product of doublet
levels at each site. The operator $\hat{T}_{ij}$ connects these
states to intermediate many-body levels $(d^6,d^8)$ or
$(d^8,d^6)$ which we denote as $\ket{e}$.
(iv) From second order perturbation theory,
we obtain the effective Hamiltonian as
\bea
\left[\mathcal{H}^{(2)}_{\rm eff}\right]_{b',b}
= \sum_e { \bra{b'} \hat{T}_{ij} \ket{e} \bra{e} \hat{T}_{ij}
\ket{b} \over E_b - E_e }
\eea
where $E_e$ and $E_b=E_{b'}$ refer to the energies of the intermediate and initial/final ground states, respectively.
$\mathcal{H}^{(2)}_{\rm eff}$ is a 4x4 matrix, describing the interactions between the doublet manifolds on the two sites.
(v) Finally, from this 4x4 Hamiltonian, the exchange matrix $\mathcal{J}$ is extracted by writing $\mathcal{H}^{(2)}_{\rm eff}$ above as a pseudospin exchange term
    \bea
    \mathcal{H}^{ij} = \sum_{\alpha\beta} \; {\mathcal S}^i_\alpha \; \mathcal{J}^{ij}_{\alpha \beta} \; {\mathcal S}^j_\beta
    \eea
    where $\alpha$ and $\beta$ are the components of the spin vectors, and the $\mathcal{S}_\alpha$ are pseudospin-$1/2$ operators.

For instance, along a $C$-type nearest-neighbor bond (see left panel in Fig.\ref{fig:spinbases}), the
exchange matrix $\mathcal{J}[C]$ is defined as
\begin{equation}
\mathcal{J}[C] =
\begin{pmatrix}
	J^{XY}+D & E & G \\
	E & J^{XY}-D & F \\
	G & F & J^Z
\end{pmatrix}
\end{equation}
with antisymmetric terms, e.g. Dzyaloshinskii-Moriya (DM) exchange
couplings, being forbidden by inversion symmetry about the
bond center.
We can extract the corresponding exchange matrices $\mathcal{J}[A]$ and
$\mathcal{J}[B]$ on the $A$-type and $B$-type bonds
simply by a $\pm 2\pi/3$ rotations
of the above exchange matrix about the $C_3$ axis. More generally, the exchange matrices on generic bonds would allow DM terms,
but we find that these are negligible, even for bonds
where there is no center of inversion (eg. second-neighbor bonds $2$ and $2'$).

In Appendix \ref{app_kitaevbasis}, we present results for the exchange matrices in terms of the ideal local octahedral basis in which the famous Kitaev-$\Gamma$ type Hamiltonians are usually formulated (also shown in the left panel in Fig.\ref{fig:spinbases}). However, we note that in all the cobaltates
we have examined, the pseudospin Hamiltonian is simplest in the above global basis, i.e. with the fewest number of nonzero entries in the exchange matrices.

\subsection{Dependence of Exchanges on Interaction Parameters}

\begin{figure}[t]
  \centering
    \centering
    \includegraphics[width=0.48\textwidth]{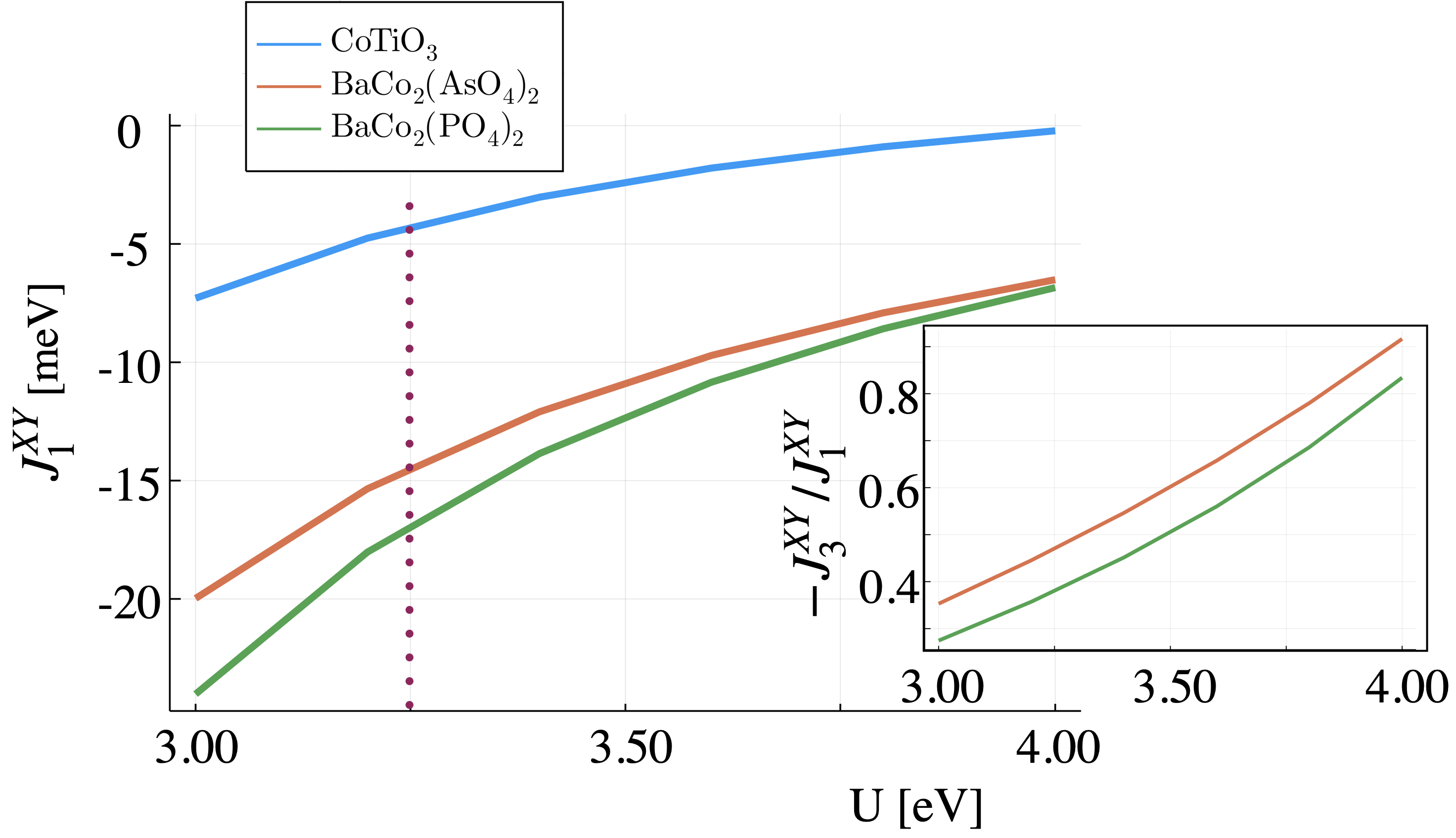}
    \caption{Dependence of the nearest-neighbor exchange interaction $J_1^{XY}$ on the Hubbard $U$, for fixed $J_H=0.7$ eV.
    For $U=3.25$ eV (shown by the dotted line), the exchange couplings for CoTiO$_3$ are in good agreement with neutron experiments
    \cite{Yuan2020,Elliot2020}.
    Inset shows the ratio $-J_3^{XY}/J_1^{XY}$ over the same range of $U$ for the Barium compounds, suggesting significant frustration from third
    neighbor exchange. The full set of significant exchange couplings for $U=3.25$ and $J_H=0.7$\,eV are quoted in Tables \ref{table:ctoex}, \ref{table:bcaoex}, and \ref{table:bcpoex}.}
  \label{fig:Udep}
\end{figure}

The exchange interactions
are sensitive to the interaction parameters
$U,J_H$ in Equation (\ref{interactionham}). To illustrate this,
Fig.~\ref{fig:Udep} shows the dependence of the
first-neighbor $XY$ exchange $J^{XY}_1$
on the Hubbard repulsion $U$ for all three cobaltates,
while the inset displays the
variation of the non-negligible third-neighbor exchange $-J^{XY}_3/J^{XY}_1$
in BaCo$_2$(AsO$_4$)$_2$ and BaCo$_2$(PO$_4$)$_2$.
The SM \cite{SuppMat} includes
additional plots for the $J_H$ dependence of these couplings, as well as the ($U,J_H$) dependence of bond-dependent
anisotropies and further neighbor interactions.

Based on Fig.~\ref{fig:Udep}, and similar plots, we find that a reasonable
choice of $U=3.25$\,eV
and $J_H=0.7$\,eV \cite{WangU2006,Cortada2016} leads to exchange couplings for CoTiO$_3$ which are in excellent agreement with inelastic neutron scattering experiments.
For BaCo$_2$(AsO$_4$)$_2$ and BaCo$_2$(PO$_4$)$_2$, however,
the spin ground state is extremely sensitive
to the third-neighbor exchange over a range of $J^{XY}_3/J^{XY}_1$ (shown in the inset of Fig.~\ref{fig:Udep}). This suggests that
some degree of fine-tuning might be needed to explain their
precise magnetic ordering. Indeed, experiments on
BaCo$_2$(PO$_4$)$_2$ have even argued for the coexistence of orderings with
different wavevectors in the same sample \cite{Nair2018}. In addition to this
issue,
we find a much larger overall scale of the exchange couplings
in these
two materials when compared with experimental reports - we
later discuss a possible reason for this significant discrepancy.

\section{Spin Hamiltonians, ordered states, spin dynamics}

\subsection{CoTiO$_3$}
We find that the spin model in CoTiO$_3$ is best described as a nearest-neighbor ferromagnetic easy-plane $XXZ$ model with significant compass-type anisotropies in the honeycomb plane, along with a weaker antiferromagnetic coupling between honeycomb layers. The full set of significant exchange coupling values are listed in Table \ref{table:ctoex}. This complete set of spin interactions favors an ordered
ground state
where the moments have $XY$ ferromagnetic order within each honeycomb layer, but
are antiferromagnetically stacked from one layer to the next.
The interlayer exchanges are mediated by the significant
overlap of the Co Wannier orbitals with the spacer Ti ions.
The first-neighbor exchange coupling and its strong anisotropy, as well as the
weaker interlayer couplings, are in very good
agreement with those inferred from neutron scattering studies
of the material \cite{Yuan2020,Elliot2020}. The scale of the compass
anisotropy terms are consistent with those conjectured
in previous work \cite{Elliot2020}.


\begin{table}[t]
\centering
\begin{tabular}{|P{2.1cm}|P{2cm}|P{2cm}|P{2cm}|}
	\hline
	\textbf{CoTiO$_3$} & Bond-\textbf{1} & Bond-\textbf{1'} & Bond-\textbf{2'} \\
	\hline\hline
	$J^{XY}$ & $-4.5$\,meV & $0.4$\,meV & $0.8$\,meV \\
	\hline
	$J^Z$ & $-0.3$\,meV & $0.3$\,meV & $0.4$\,meV \\
	\hline
	$D$ & $1.8$\,meV & $\sim 0$ & $\sim 0$ \\
	\hline
	$E$ & $-2.2$\,meV & $\sim 0$ & $\sim 0$ \\
	\hline
\end{tabular}
\caption{Table of dominant exchange parameters for CoTiO$_3$, corresponding to $U=3.25$ eV and $J_H=0.7$ eV. For the definitions of the bonds, see Fig.~\ref{fig:spinbases} (right panel).}
\label{table:ctoex}
\end{table}

\begin{figure}[t]
\includegraphics[width=0.48\textwidth]{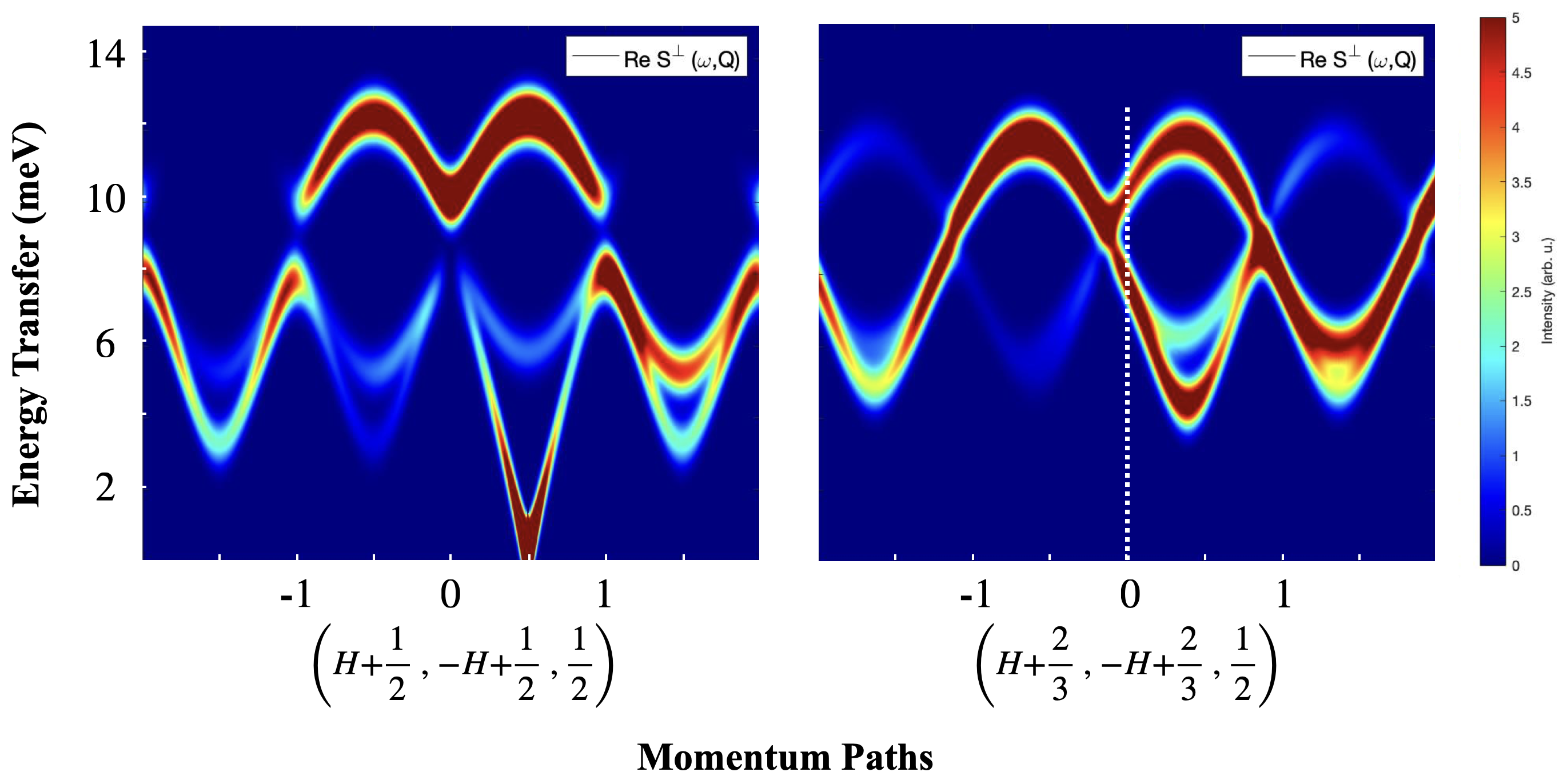}
\caption{Spin wave dispersions for CoTiO$_3$, obtained using SpinW \cite{Toth2014}, plotted along two cuts in
momentum space in the $L=1/2$ plane \cite{Yuan2020,Elliot2020}.
Dotted line in right panel denotes the Brillouin zone corner.}
\label{fig:ctosw}
\end{figure}

Fig.~\ref{fig:ctosw} shows the linear spin wave dispersion, obtained with our
computed exchange interactions, along specific cuts in momentum space. The left panel reveals a
Goldstone mode at the zone center - this is an artefact of linear spin wave theory, since
the significant anisotropic ``compass''-type exchange couplings $D$ and $E$ break the $U(1)$
symmetry of the pure $XXZ$ model. Order by disorder physics, beyond linear spin-wave theory,
is expected to pin the ordered moment and gap out the Goldstone mode \cite{Elliot2020}. The right panel shows
the computed Dirac node which has been observed in
inelastic neutron scattering experiments; we find that this
node is shifted away from the Brillouin zone corner
which is at $(2/3,2/3,1/2)$ (dotted line in the right panel of Figure \ref{fig:ctosw}). This is in accordance with
predictions that
there are 3D Dirac nodal lines which should wind, like a
triple helix, around the $(2/3,2/3,L)$ axis in momentum space. This displacement of the
Dirac nodal lines is enabled by the anisotropic $D$ and $E$
exchange couplings, and might lead to a suppression of
the magnon density of states, as seen in recent experiments \cite{Elliot2020}.

\subsection{BaCo$_2$(AsO$_4$)$_2$ and BaCo$_2$(PO$_4$)$_2$}
The Barium compounds are much more
two-dimensional compared to CoTiO$_3$, with extremely small
inter-layer couplings
$\lesssim 0.1$\,meV. The
source of this layer decoupling is the large inter-plane
distances discussed in Section \ref{sec:crystalstructure}.
In addition, the less buckled nature of the honeycomb planes
and the Wannier orbital overlap with As/P appears to promote
larger in-plane hopping amplitudes. As a result, we find that
the computed exchange interactions between $j\!=\! 1/2$ pseudospins
in these compounds are significantly
enhanced compared with CoTiO$_3$.

\begin{table}[t]
\centering
\begin{tabular}{|P{2.3cm}|P{2cm}|P{2cm}|P{2cm}|}
	\hline
	\textbf{BaCo$_2$(AsO$_4$)$_2$} & Bond-\textbf{1} & Bond-\textbf{3} \\
	\hline\hline
	$J^{XY}$ & $-14.5$\,meV & $7.1$\,meV \\
	\hline
	$J^Z$ & $-3.8$\,meV & $2.3$\,meV \\
	\hline
	$D$ & $1.5$\,meV & $\sim 0$ \\
	\hline
	$E$ & $-1.7$\,meV & $\sim 0$ \\
	\hline
\end{tabular}
\caption{Table of dominant exchange parameters for BaCo$_2$(AsO$_4$)$_2$, corresponding to $U=3.25$ eV
and $J_H=0.7$ eV. For the definitions of the bonds, see Fig.~\ref{fig:spinbases} (right panel).}
\label{table:bcaoex}
\end{table}

\begin{table}[b]
\centering
\begin{tabular}{|P{2.3cm}|P{2cm}|P{2cm}|P{2cm}|}
	\hline
	\textbf{BaCo$_2$(PO$_4$)$_2$} & Bond-\textbf{1} & Bond-\textbf{3} \\
	\hline\hline
	$J^{XY}$ & $-17.3$\,meV & $6.9$\,meV \\
	\hline
	$J^Z$ & $-5.0$\,meV & $2.4$\,meV \\
	\hline
	$D$ & $-2.7$\,meV & $\sim 0$ \\
	\hline
	$E$ &  $0.2$\,meV & $\sim 0$ \\
	\hline
\end{tabular}
\caption{Table of dominant exchange parameters for BaCo$_2$(PO$_4$)$_2$, corresponding to $U=3.25$ eV
and $J_H=0.7$ eV. For the definitions of the bonds, see Fig.~\ref{fig:spinbases} (right panel).}
\label{table:bcpoex}
\end{table}

Tables \ref{table:bcaoex} and \ref{table:bcpoex} show values for
the exchange parameters for the BaCo$_2$(AsO$_4$)$_2$ and BaCo$_2$(PO$_4$)$_2$ respectively.
The first-neighbor interaction leads to an easy-plane $XXZ$
model, but with subdominant anisotropies, which is qualitatively similar to CoTiO$_3$. However, in contrast to CoTiO$_3$
we find strong
frustration coming from significant third-neighbor
antiferromagnetic exchange; this leads to
antiferromagnetic $XY$ ground
states in the Barium compounds.

The classical phase diagram of
the $J_1$-$J_3$ model   \cite{RASTELLI1979,J.B.FouetP.SindzingreC.Lhuillier2004}
features a tiny
sliver ($0.25 \!\lesssim\! -J_3/J_1 \! \lesssim 0.4$) of an
incommensurate coplanar spiral phase sandwiched
 between the more robust ferromagnetic and
zig-zag ordered phases; see Appendix B
for a phase diagram reproduced from Ref.~\cite{J.B.FouetP.SindzingreC.Lhuillier2004}.
Based on our exploration of how the
exchange couplings vary with the interaction parameters in
Eq.~\ref{interactionham} (see Figure \ref{fig:Udep}), we find
that the ratio $J_3/J_1$ lies close to this window. We thus
deduce that the ground states of these materials could be
extremely sensitive to small variations in material
parameters.


Based on our computed exchange couplings shown in Tables
\ref{table:bcaoex} and \ref{table:bcpoex}, we extract
the Curie-Weiss constants for BaCo$_2$(AsO$_4$)$_2$ and BaCo$_2$(PO$_4$)$_2$ both parallel to the $c$-axis,
and perpendicular to
the $c$-axis (i.e., in the honeycomb plane). For BaCo$_2$(AsO$_4$)$_2$, we find
$\Theta_\parallel \!\simeq \! 65$\,K and
$\Theta_\perp \!\simeq \! 4$\,K, about twice as large as
the corresponding experimental values \cite{Zhong2020}. For BaCo$_2$(PO$_4$)$_2$,
we find $\Theta_\parallel \!\simeq \! 90$\,K and
$\Theta_\perp \!\simeq \! 23$\,K, which remains to be
experimentally studied on single crystal samples.

\begin{figure}[t]
\includegraphics[width=0.48\textwidth]{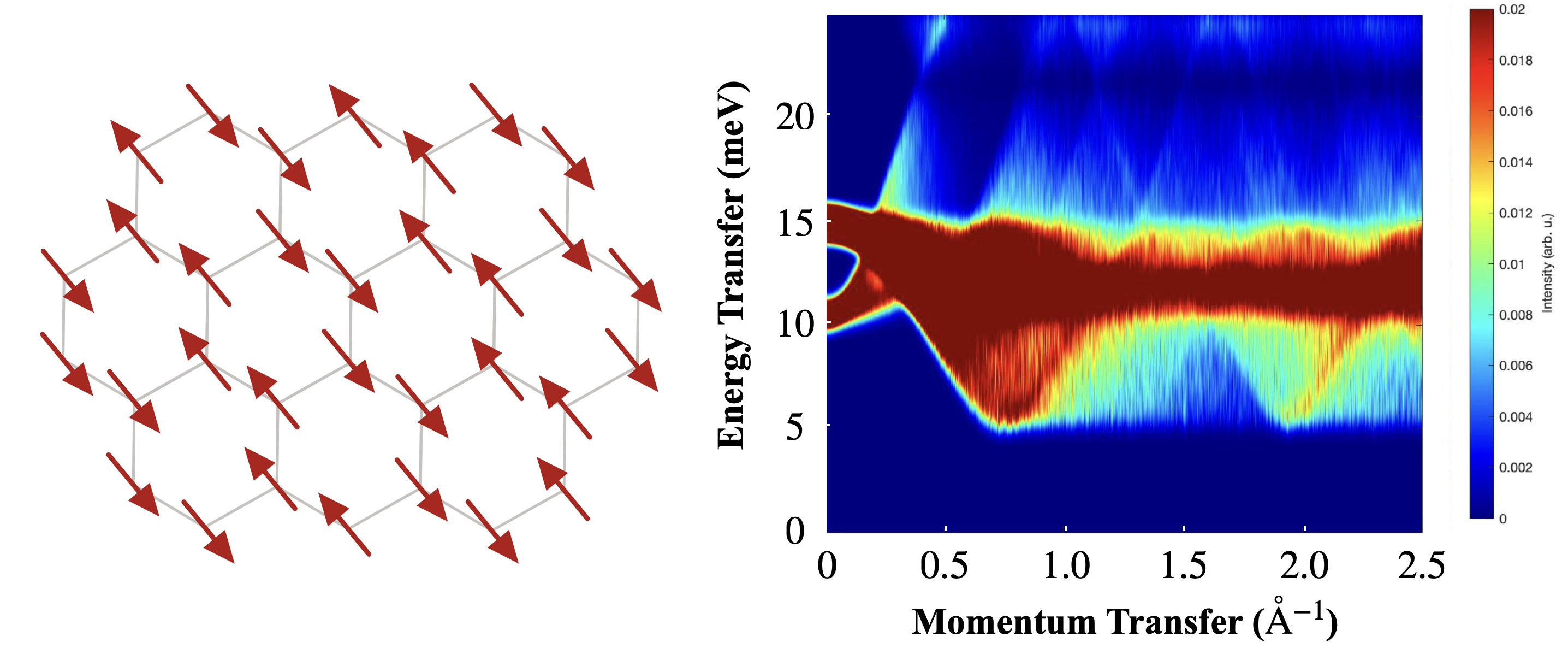}
\caption{Left panel: Zig-zag ordered phase with $\bQ_0 \!=\! (1/2,0)\!\approx\!(0.628,0)$\AA$^{-1}$.
Right panel: Powder averaged dynamical structure factor of the zig-zag ordered phase for the parameter set from
Table \ref{table:bcaoex}.}
\label{fig:mfig}
\end{figure}

\begin{figure}[b]
\includegraphics[width=0.5\textwidth]{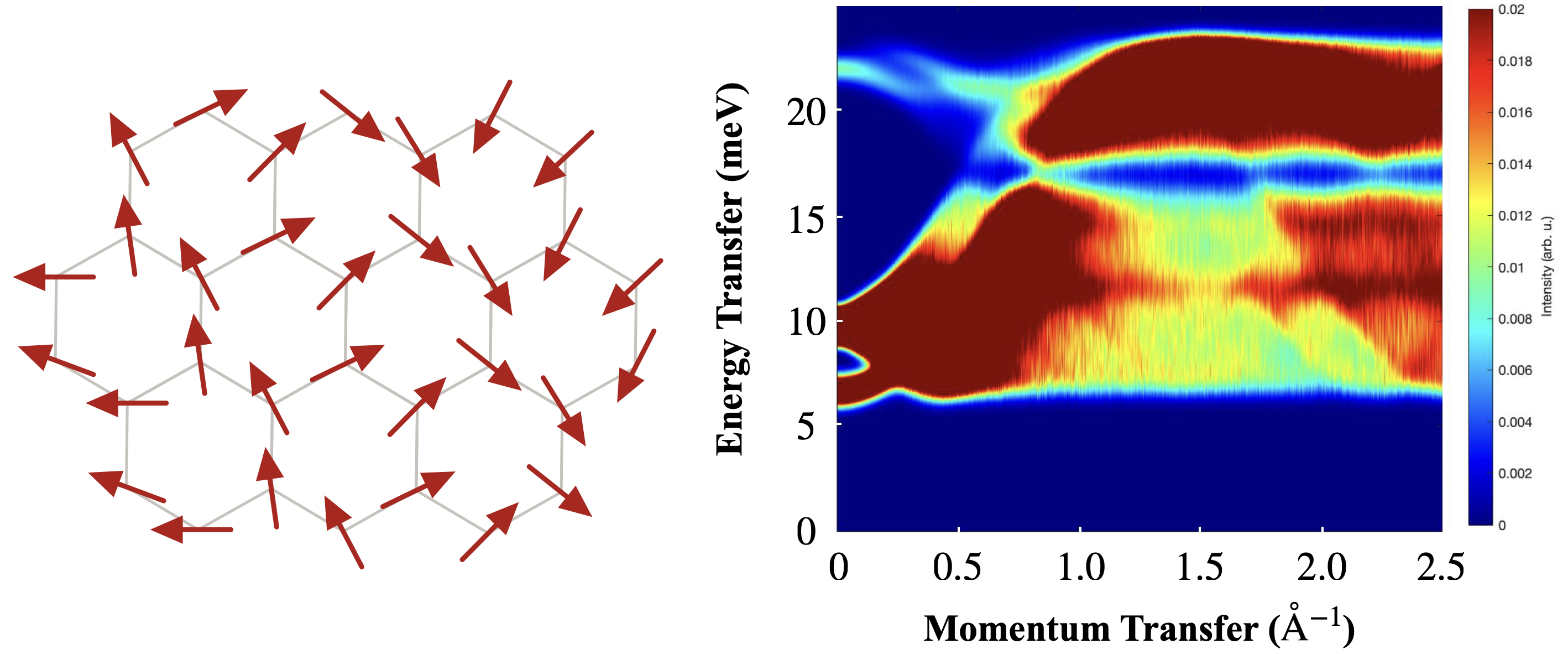}
\caption{Left panel:
Spiral ordered phase with $\bQ_0 = (0.34,0)${\AA}.
Right panel: Powder averaged dynamical structure factor of the zig-zag ordered phase for the parameter set from
Table \ref{table:bcaoex} but with all third-neighbor couplings rescaled such that $J^{XY}_3/J^{XY}_1 \!=\! -0.3$.}
\label{fig:mb2fig}
\end{figure}

Using SpinW \cite{Toth2014}, we have computed the magnetic order and
dynamical spin correlations
corresponding to the exchange values in Table
\ref{table:bcaoex}. We find that this parameter set
supports collinear zig-zag magnetic order as shown in Fig.~\ref{fig:mfig}
(left panel), with moment directions as shown for the ordering
wavevector along $\hat{X}$. The wavevector $(H,K)$ for this order
corresponds to $\bQ_0=(1/2,0)$;
reinstating the lattice units,
this translates to $\bQ_0\!\simeq (0.628,0)$\AA$^{-1}$.
The dynamical spin correlations
corresponding to this order show significant intensity
at a band of
excitations around an energy transfer
$\hbar\omega\!\sim\! J^{XY}_1$. In addition, the Goldstone
mode expected at $\bQ_0$ for a simple $XXZ$ model is gapped out by
the anisotropy terms in the Hamiltonian, with a gap
$\sim\!0.3 J^{XY}_1$.
Below the scale of this Goldstone gap, we expect that
the magnetic specific heat should be exponentially
suppressed. Similar results follow for the exchange parameter
set from Table \ref{table:bcpoex}.

On the other hand, experiments on BaCo$_2$(AsO$_4$)$_2$ and BaCo$_2$(PO$_4$)$_2$ do not find a
zig-zag ordered ground state. While early experiments on BaCo$_2$(AsO$_4$)  \cite{Regnault1977} reported
results consistent with a coplanar spiral (helical) ground state with a wavevector
$\bQ_0 \! \simeq \! (0.27,0)$, or equivalently $(0.34,0)$\AA$^{-1}$, recent work
\cite{Regnault2018} has instead suggested a collinear order with this wavevector.
For BaCo$_2$(PO$_4$), recent experiments \cite{Nair2018} have argued for short-range ordering involving
both zig-zag and spiral states.
Overall, these experiments suggest a slightly smaller third-neighbor coupling, which would favor a smaller
wavevector compared with the zig-zag ordered ground state,
reflecting the extreme sensitivity of the magnetic ground state
to small changes in the frustration. This, in turn, might arise
from small differences in the microscopic Kanamori interactions.
To illustrate this, we have computed the magnetic ground state
and dynamical spin correlations by rescaling all third-neighbor
hoppings by a factor $\approx\!0.8$, which leads to
a renormalized $J_3^{XY}/J^{XY}_1 \!=\! -0.3$. In this
case, we find that the magnetic ground state has the
experimentally reported ordering wavevector, with the spin
order as shown in Fig.~\ref{fig:mb2fig} (left panel). The
corresponding dynamical spin correlations are shown in
Fig.~\ref{fig:mb2fig} (right panel). As expected, the
Goldstone mode continues
to be gapped due to the anisotropic exchange terms.
Indeed, inelastic neutron scattering experiments on BaCo$_2$(AsO$_4$)$_2$ and BaCo$_2$(PO$_4$)$_2$ find
evidence for spin gapped excitations, qualitatively consistent
with our results \cite{Regnault2018,Nair2018}.

It is possible that the collinear order
reported in \cite{Regnault2018} could reflect the impact of strong quantum fluctuations on the competition between
spiral and collinear ground states. Such competition has been extensively investigated in spatially anisotropic triangular
antiferromagnets \cite{StarykhPRL2007};
we defer a full quantum analysis of our model to a future publication.

Specific heat measurements on BaCo$_2$(AsO$_4$)$_2$ have
reported a $\sim\!T^2$ specific
heat at the lowest temperature $T \!\lesssim 2$\,K
 \cite{BCAO1978,Regnault2018}, which was attributed to 2D magnetic
 fluctuations. However, our calculations suggest that the
specific heat should be suppressed below a spin gap scale $\sim\! 0.3 J_1^{XY}$; we thus conclude that the experimentally observed
 specific heat may be of non-magnetic
origin. Alternatively, it might reflect the impact of disorder
which has not been taken into account in our work.

One key discrepancy between our calculations
and the experimental results on BaCo$_2$(AsO$_4$)$_2$ and BaCo$_2$(PO$_4$)$_2$ is in the overall scale of the
magnetic exchange.
The exchange couplings we find are about $3-4$ times larger than those reported in the experimental literature.
We have explored varying $U,J_H$ and the crystal field matrix in order to resolve this discrepancy while maintaining
a reasonable ratio of $J^{XY}_3/J^{XY}_1$, but without
any success. We therefore propose the following tentative explanation to reconcile this problem. We have
seen that the Co$^{2+}$ ion supports a lowest energy single-ion excitation, i.e. a spin exciton,
at an energy $\simeq 20$-$30$\,meV. In CoTiO$_3$, the
exchange interactions between the $J_{\rm eff}\!=\!1/2$ pseudospins is
much smaller than this spin-exciton gap, justifying the
restriction to a pseudospin-$1/2$ model. By contrast,
the magnetic exchange interactions in BaCo$_2$(AsO$_4$)$_2$ and BaCo$_2$(PO$_4$)$_2$ are comparable to the spin-exciton gap.
This suggests that the spin dynamics and Curie-Weiss
temperatures might both be strongly impacted by coupling between
the low energy magnon and the spin exciton. Level repulsion between these states
could then lead to an effectively smaller magnon bandwidth for the $J_{\rm eff}\!=\! 1/2$ spins.
The theory of this magnon-exciton
coupling will be discussed elsewhere.

Our viewpoint taken in this paper is that we can focus on a Co-only model with site-local interactions,  although Fig.~3 shows that the Wannier orbitals do have
oxygen character. To what extent is this a reasonable
assumption? Partial justification for retaining only an effective Co interacting model stems from the recognition that compounds in the transition metal oxide series go
from being Mott-Hubbard insulators to charge-transfer insulators as we go from Ti to Cu, with the cuprates being well-known charge transfer insulators. Recent work suggests that nickelates, such as NdNiO2, lie at the borderline \cite{Goodgee2007683118}. Based on this, the cobaltates, are perhaps better viewed as Mott-Hubbard insulators
where focussing on excitations on the Co d-levels is a good starting point, and indeed we do seem to find excellent agreement with experiments on CoTiO$_3$
following this approach. Nevertheless, in light of the above discrepancies, a more quantitative treatment of the two-site exchange interaction incorporating
oxygen orbitals, in addition to the Co ions, might might be important to consider in future studies of BaCo$_2$(AsO$_4$)$_2$ and BaCo$_2$(PO$_4$)$_2$.

\section{Discussion}

We have explored the magnetism of the honeycomb
cobaltates CoTiO$_3$, BaCo$_2$(AsO$_4$)$_2$, and BaCo$_2$(PO$_4$)$_2$
which crystallize in the rhombohedral
structure with $R\bar{3}$ space group. The Co$^{2+}$ ions in the
$d^7$ configuration have been proposed as possible
candidate materials
for realizing strong Kitaev interactions and Kitaev quantum
spin liquids. Based on our combined DFT and exact diagonalization
study, we find that the magnetism in these cobaltates
is strongly perturbed by the trigonal distortion,
and that they
are better described as $XXZ$ quantum magnets with subdominant
but nevertheless important anisotropy terms. Our results for CoTiO$_3$ are
in excellent agreement with neutron scattering experiments \cite{Yuan2020,Elliot2020}.
The bond-anisotropic exchanges we have uncovered
can gap the Goldstone modes in these magnets, and lead
to a helical winding of the Dirac magnon nodal line in
CoTiO$_3$. In BaCo$_2$(AsO$_4$)$_2$ and BaCo$_2$(PO$_4$)$_2$, we
have found microscopic evidence for strong third-neighbor antiferromagnetic
exchange, which can drive possibly quite distinct quantum spin liquid phases
proximate to zig-zag or spiral ordered phases. Indeed, the reported collinear order \cite{Regnault2018}
and small magnetic field to drive saturation \cite{Zhong2020} point to the importance
of strong quantum fluctuations in these two materials.
In addition, we have presented arguments that the low-energy spin excitations
in the Ba compounds may involve strongly coupled magnons and spin excitons.
While such magnon-exciton coupling has been explored
within phenomenological models for CoTiO$_3$ \cite{Elliot2020}
and FeI$_2$ \cite{Bai2021}, it would be
worth revisiting those ideas in light of our work in order to provide a quantitative
understanding of BaCo$_2$(AsO$_4$)$_2$ and BaCo$_2$(PO$_4$)$_2$,
and their magnetic field driven phase diagram \cite{Zhong2020,Armitage2021,Shi2021}.
Carefully integrating out the high energy exciton may lead to an enhancement of the Kitaev
exchange interaction for the $J_{\rm eff}\!=\!1/2$ pseudospins; we are presently exploring this possibility.

While we have explored a
limited set of all the honeycomb cobaltates, i.e. the rhombohedral cases,
most other candidate cobaltates have even lower symmetry.
For instance,
Na$_3$Co$_2$SbO$_6$ \cite{Viciu2007,Songvilay2020,Lefrancois2016,Xiao2019,Bera2017,Wong2016}, Ag$_3$Co$_2$SbO$_6$ \cite{Zvereva2016}, Li$_3$Co$_2$SbO$_6$ \cite{Brown2019,Stratan2019},
and CoPS$_3$ \cite{Wildes2017} crystallize in the monoclinic $C2/m$ space group,
while Co$_4$Nb$_2$O$_9$ and Co$_4$Ta$_2$O$_9$ \cite{Bertaut1961,Khanh2016,Yanagi2018a,Yanagi2018}
crystallize in the $P\bar{3}c1$ space group.
 We thus suspect that they may support even more complex magnetic exchange interactions.
 In addition, the zig-zag magnetic order
observed in many of these compounds may again stem from a significant third-neighbor coupling as was initially proposed
also for the iridates \cite{Kimchi2011} and also recognized to be important for the cobaltates \cite{Liu2018}.

As a test case, we have carried out a preliminary study of the exchange couplings in Na$_3$Co$_2$SbO$_6$.
We find that its spin Hamiltonian is consistent with nearly decoupled 2D honeycomb layers.
However, the $3\times 3$ matrix of in-plane exchange couplings does not appear to be simple (i.e., sparse) in either the local octahedral basis or
a global spin basis; indeed all couplings permitted by symmetry appear to be relevant.
Our preliminary calculations yield, in the local octahedral basis, a nearest-neighbor ferromagnetic Heisenberg coupling $J_1 \!\sim\! -3$\,meV,
a comparable antiferromagnetic Kitaev exchange $K \sim\! 2.5$\,meV, an off-diagonal symmetric exchange
$\Gamma \!\sim\! -2.5$\,meV, and a frustrating antiferromagnetic third-neighbor Heisenberg exchange $J_3 \! \sim \! 2.5$\,meV,
together with additional bond-anisotropic  terms $\Gamma'_1,\Gamma'_2$. The significant $J_3$ appears to be mediated by the Sb ion in the honeycomb
plane. This spin Hamiltonian favors a zig-zag ground state with a gapped Goldstone
mode, in qualitative agreement with neutron diffraction and inelastic scattering experiments
\cite{SongvilayPRB2020,kim2020antiferromagnetic}. Our estimated
exchange couplings are roughly in the same ballpark as those extracted from careful spin-wave fits to the inelastic neutron scattering data
\cite{SongvilayPRB2020,kim2020antiferromagnetic}, but differ in detail (e.g., the sign of $\Gamma$).
Clearly, given the absence of $C_3$ symmetry in this monoclinic crystal, a more careful study is needed to reliably extract the
exchange interactions and model the spin-wave spectrum. Nevertheless, our preliminary calculations appear to confirm that the spin model is far more
complex in this lower symmetry magnet, suggesting that this material also does not realize a simple
Kitaev-exchange dominated spin model. The presence of multiple competing interactions could still lead to
nearby quantum spin liquid phases whose existence and nature remain open issues.

In other cobaltates, such as the recently studied Na$_2$Co$_2$TeO$_6$ \cite{samarakoon2021static} which crystallizes in the
$P6_3 22$ space group and exhibits zig-zag magnetic order,
randomness in the Na$^+$ arrangement might promote disorder-enabled spin liquid signatures. However, this randomness
stymies a full {\it ab initio} based study of the exchange interactions. Progress may be made by exploring a suitable periodic
arrangement of Na$^+$ ions in a supercell; we defer this to a future investigation.

In summary, we argue
that the honeycomb cobaltates can realize a rich set of magnetic
Hamiltonians with frustration and anisotropic bond-dependent interactions. and may potentially host interesting
quantum spin liquids driven by further-neighbor interactions and disorder. Realizing the Kitaev quantum spin liquid in this class of materials
may require further fine-tuning of the chemistry and
crystal fields in order to promote the Kitaev coupling over other competing exchange interactions.

\acknowledgements
We thank Giniyat Khaliullin, Kate Ross, Peter Armitage, and Sasha Chernyshev for useful discussions.
SV and AP acknowledge funding from NSERC of Canada.
TS-D acknowledges a J.C. Bose National Fellowship (grant no. JCB/2020/000004) for funding. SD was supported by a DST-INSPIRE fellowship.
Exact diagonalization computations were performed on the Niagara
supercomputer at the SciNet HPC Consortium. SciNet is funded by: the Canada Foundation for Innovation; the
Government of Ontario; Ontario Research Fund - Research Excellence; and the University of Toronto.

\bibliography{d7exchange.bib}

\appendix

\section{Exchange couplings in local Kitaev basis}
\label{app_kitaevbasis}

In this Appendix, we discuss the basis rotation and the rewriting of the exchange couplings in the Kitaev basis.
The spins written in the global $XYZ$ basis ($\mathcal{S}_\alpha$ can be transformed into the local $\tilde{x}\tilde{y}\tilde{z}$ basis ($\tilde{\mathcal{S}}_\alpha$) via a transformation $\tilde{\mathcal{S}}_\alpha = \mathcal{U}_{\alpha\beta}\mathcal{S}_\beta$, where $\mathcal{U}$ is given by the matrix
\bea
\mathcal{U} = \begin{pmatrix}
    {1 / \sqrt{6}} && {1 / \sqrt{2}} && {1 / \sqrt{3}} \\
    {-\sqrt{2 / 3}} && {0} && {1 / \sqrt{3}} \\
    {1 / \sqrt{6}} && -{1 / \sqrt{2}} && {1 / \sqrt{3}}
\end{pmatrix}
\eea
Thus, the local exchange matrix $\mathcal{J}_{\rm loc}$ is obtained via
\bea
\mathcal{J}_{\rm loc} = \mathcal{U J U^T}
\eea
where $\mathcal{J}$ is the exchange matrix in the global basis. Along the same $C$ bond (referred to in the local basis as the $z$ bond), the Heisenberg ($J$) , Kitaev ($K$), Diagonal ($\eta$) and off-diagonal ($\Gamma$, $\Gamma_1'$, $\Gamma_2'$) anisotropy couplings are defined as
\bea
\mathcal{J}_{\rm loc}^{(z)} = \begin{pmatrix}
    J+\eta && \Gamma && \Gamma_1' \\
    \Gamma && J-\eta && \Gamma_2' \\
    \Gamma_1' && \Gamma_2' && J+K
\end{pmatrix}
\eea
Below we present the couplings in this local basis for the materials considered. It should be noted that many couplings have comparable values, which does not
allow for as simple a description as the $XXZ$ model in the global basis which is presented in the main text.

\begin{table}[H]
\centering
\begin{tabular}{|P{2.3cm}|P{2cm}|P{2cm}|P{2cm}|}
	\hline
	\textbf{CoTiO$_3$} & Bond-\textbf{1} & Bond-\textbf{1'} & Bond-\textbf{2'} \\
	\hline\hline
	$J$ & -3.80 & 0.33 & 1.07 \\
	\hline
	$K$ & 1.37 & 0 & 0.12 \\
	\hline
	$\Gamma$ & 2.08 & 0 & -0.17 \\
	\hline
	$\eta$ & -1.84 & 0 & 0 \\
	\hline
	$\Gamma_1'$ & -0.57 & 0 & -0.18 \\
	\hline
	$\Gamma_2'$ & 2.73 & 0 & -0.35 \\
	\hline
\end{tabular}
\caption{Local exchange couplings for CoTiO$_3$.}
\label{table:ctolocalex}
\end{table}

\begin{table}[H]
\centering
\begin{tabular}{|P{2.3cm}|P{2cm}|P{2cm}|}
	\hline
	\textbf{BaCo$_2$(AsO$_4$)$_2$} & Bond-\textbf{1} & Bond-\textbf{3}  \\
	\hline\hline
	$J$ & -12.40 & 5.51 \\
	\hline
	$K$ & 2.06 & -0.30 \\
	\hline
	$\Gamma$ & 4.05 & -1.44 \\
	\hline
	$\eta$ & -0.84 & 0.00 \\
	\hline
	$\Gamma_1'$ & 1.82  & -1.42 \\
	\hline
	$\Gamma_2'$ & 4.89 & -1.45 \\
	\hline
\end{tabular}
\caption{Local exchange couplings for BaCo$_2$(AsO$_4$)$_2$.}
\label{table:baolocalex}
\end{table}

\begin{table}[H]
\centering
\begin{tabular}{|P{2.3cm}|P{2cm}|P{2cm}|}
	\hline
	\textbf{BaCo$_2$(PO$_4$)$_2$} & Bond-\textbf{1} & Bond-\textbf{3} \\
	\hline
	$J$ & -13.04 & 5.21 \\
	\hline
	$K$ & -3.11 & -0.05 \\
	\hline
	$\Gamma$ & 2.00 & -1.05 \\
	\hline
	$\eta$ & -1.35 & 0.00 \\
	\hline
	$\Gamma_1'$ & 6.41 & -1.45 \\
	\hline
	$\Gamma_2'$ & 3.38 & -1.24 \\
	\hline
\end{tabular}
\caption{Local exchange couplings for BaCo$_2$(PO$_4$)$_2$.}
\label{table:bpolocalex}
\end{table}

\section{$J_1$-$J_2$-$J_3$ Phase Diagram}

The figure below shows the classical $J_1$-$J_2$-$J_3$ phase diagram explored in \cite{J.B.FouetP.SindzingreC.Lhuillier2004}. In this work, we find that the relevant part of the phase diagram is the $y$-axis, specifically where $J_3/J1 < 0$. Experiments on the Barium com- pounds appear to stabilize the shaded ”Spiral 2” phase (characterized by a nonzero angle between spins on different sublattices. While the aforementioned study does not consider off-diagonal anisotropy terms ($D$ and $E$), we find that including them retains the general structure of the phase diagram. The narrowness of this spiral phase emphasizes the sensitivity of the magnetic order to the microscopics of the systems considered.

\begin{figure}[H]
  \includegraphics[width=.48\textwidth]{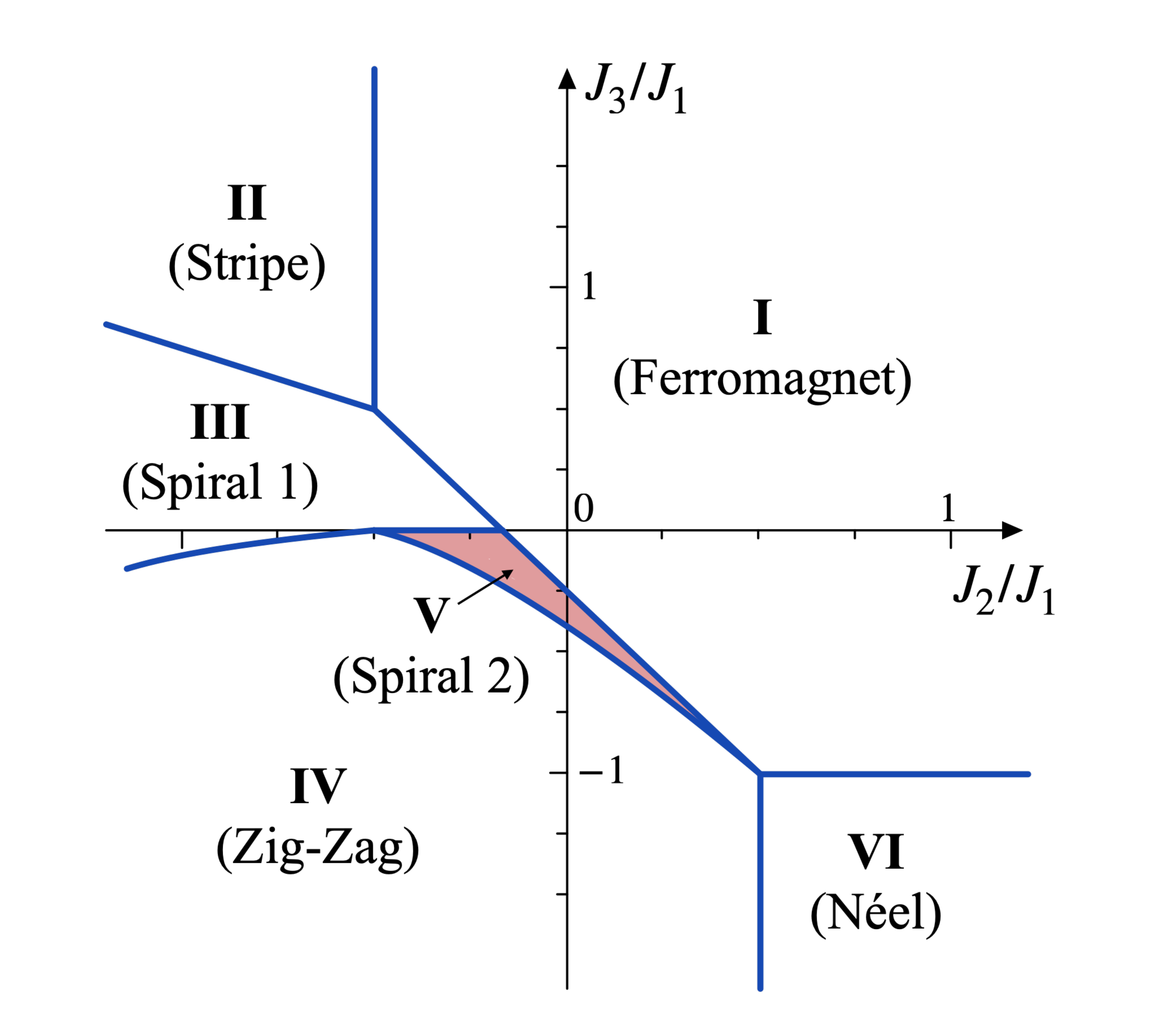}
  \caption{Classical phase diagram for the $J_1$-$J_2$-$J_3$ model, reproduced from \cite{J.B.FouetP.SindzingreC.Lhuillier2004} }
  \label{fig:j1j2j3pd}
\end{figure}

\end{document}


%

\title{SUPPLEMENTARY MATERIAL \\ $XY$ magnetism, Kitaev exchange, and long-range frustration in the $J_{\rm eff}=1/2$ honeycomb cobaltates}

\author{Shreya Das}
\thanks{These authors contributed equally}
\affiliation{Department of Condensed Matter Physics and Materials Science, S.N. Bose National Centre for Basic Sciences, Kolkata 700098, India.}

\author{Sreekar Voleti}
\thanks{These authors contributed equally}
\affiliation{Department of Physics, University of Toronto, Toronto, Ontario, M5S 1A7, Canada}

\author{Tanusri Saha-Dasgupta}
\thanks{tanusri@bose.res.in}
\affiliation{Department of Condensed Matter Physics and Materials Science, S.N. Bose National Centre for Basic Sciences, Kolkata 700098, India.}

\author{Arun Paramekanti}
\thanks{arunp@physics.utoronto.ca}
\affiliation{Department of Physics, University of Toronto, Toronto, Ontario, M5S 1A7, Canada}

\maketitle

\section{DFT Calculation Details }

 	We carried out theoretical calculations within the
 	framework of first-principles density functional theory
 	(DFT) taking into account the structural as
 	well as chemical information completely.

 	\indent Our DFT based calculations were executed with the choices of different basis sets as, (i) the plane-wave based basis as implemented in the Vienna $ab$ - $initio$ Simulation Package (VASP) \cite{vasp} with projector-augmented wave (PAW)
 	potential \cite{PAW} and (ii) the muffin-tin orbital (MTO) based linear muffin-tin orbital (LMTO) method \cite{LMTO}, and the $N^{th}$ order MTO method namely, NMTO method \cite{NMTO}, as implemented in the STUTTGART code.\\
The consistency of results between two different basis sets have been confirmed in terms density of states, band structure and magnetic moments. The exchange correlation functional was chosen as the generalized gradient approximation
 	(GGA) implemented following the Perdew-Burke-Ernzerhof (PBE) prescription \cite{PBE}.

For the plane wave calculation, energy cutoff of 600 eV and
 	MonkhorstPack k-points mesh of 6$\times$6$\times$2 for the unit cell of CoTiO$_3$ and BaCo$_2$(AsO$_4$)$_2$ were found to provide good convergence of the total energy in self-consistent field calculations. The effect of spin-orbit coupling (SOC) was included within GGA+SOC implementation of VASP and the effect of missing correlation effect beyond GGA at
 	the Co site was considered within the GGA+SOC+U frame work \cite{GGAUSOC} with the choice of $U$ = 3 eV and Hund's coupling $J_H = 0.7$ eV \cite{WangU2006,Cortada2016}.

For construction of low energy Hamiltonian, we used the $N^{th}$ order muffin-tin orbital method, which relies on a self-consistent potential generated by the
linear MTO (LMTO) method. The low-energy tight-binding Hamiltonian, defined in the effective Co Wannier basis, provides information about the crystal-field splitting at the Co site along with the effective hopping interactions between the Co
sites. The muffin-tin radii of different atomic sites used in LMTO calculations were chosen as 1.52 {\AA} for Co, 1.27 {\AA} for Ti, 0.98 {\AA} for O atoms in CoTiO$_3$; 2.44 {\AA} for Ba, 1.52 {\AA} for Co, 1.09 {\AA} for As and 0.82/0.89 {\AA} for the O atoms in BaCo$_2$(AsO$_4$)$_2$. For BaCo$_2$(PO$_4$)$_2$, the atomic radii are chosen to be 2.43 {\AA} for Ba, 1.58 {\AA} for Co, 0.97 {\AA} for P and 0.74/0.83 {\AA} for the O atoms.

\section{GGA, GGA+SOC, GGA+SOC+$U$ magnetic moments and electronic structure}

\begin {table} [h!]
\begin {tabular} {|c|ccc|cccc|}
\hline
            &      & CoTiO$_3$ &        &      & BaCo$_2$(AsO$_4$)$_2$ &      & \\ \hline
            & Co   & Ti        & O      & Ba   & Co                   & As   & O  	\\
GGA         & 2.55 & 0.11      & 0.10   & 0.00 & 2.58                 & 0.02 & 0.09     \\
GGA+SOC     & 2.55 & 0.11      & 0.10   & 0.00 & 2.58                 & 0.02 & 0.10     \\
            & (0.17) & (0.00)  & (0.00) & (0.00) & (0.15)             & (0.00) & (0.00) \\
GGA+SOC+$U$ & 2.69 & 0.10      & 0.08   & 0.00  & 2.70                & 0.02  & 0.06 \\
            & (0.19) & (0.00)  & (0.00) & (0.00) & (0.17)             & (0.00) & (0.00) \\

\hline
\end{tabular}
\caption{The magnetic moments (in $\mu_B$) at different atomic sites of CoTiO$_3$ and  BaCo$_2$(AsO$_4$)$_2$, as calculated
within GGA, GGA+SOC and GGA+SOC+$U$. The numbers in the bracket refer to orbital moments.}
\end{table}

\begin{figure}[H]
  \centering
  \includegraphics[width=.5\textwidth]{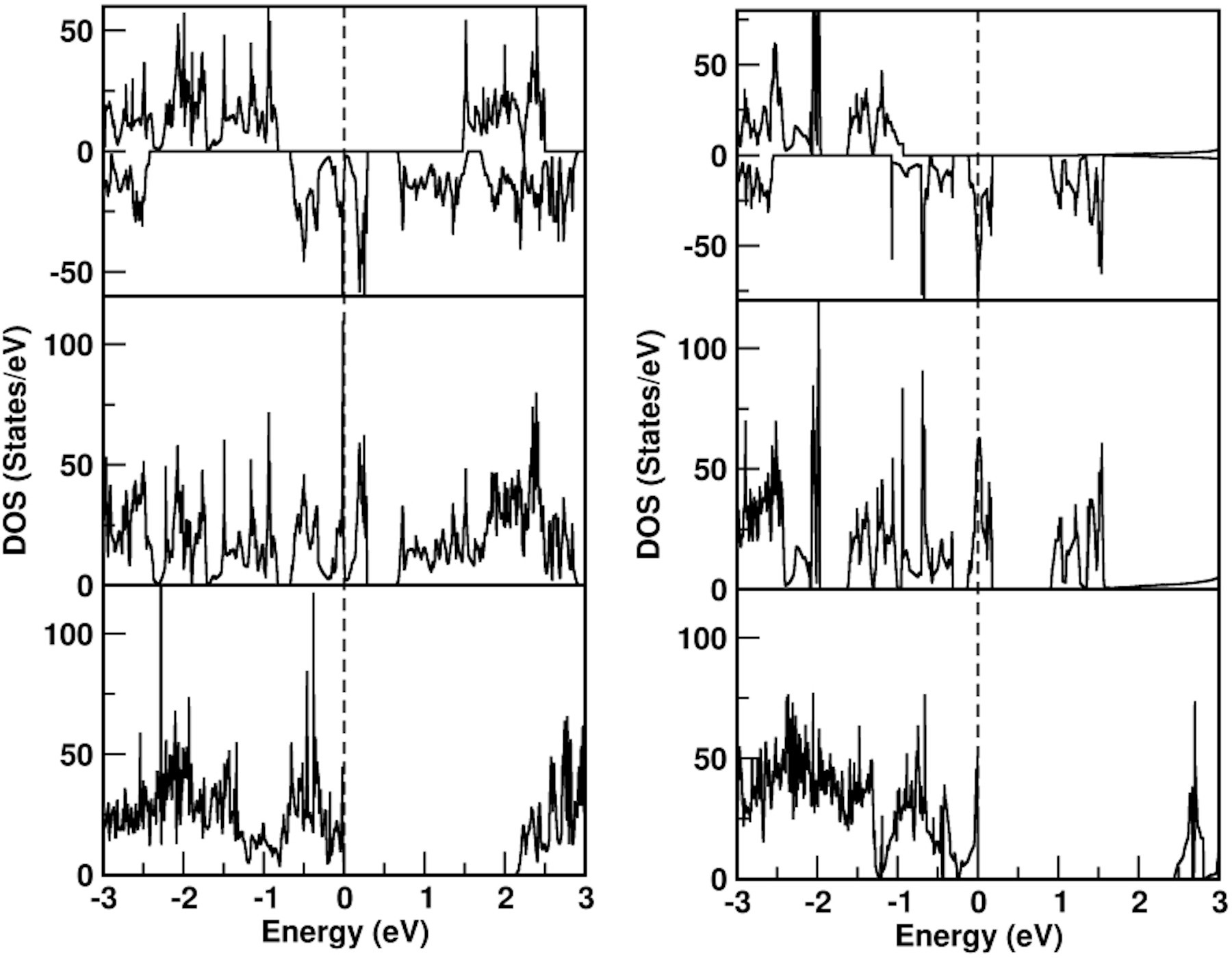}
  \caption{Density of states of CoTiO$_3$ (left panels) and  BaCo$_2$(AsO$_4$)$_2$ (right panels), as calculated
within GGA (top), GGA+SOC (middle) and GGA+SOC+$U$ (bottom). The GGA+SOC+$U$ shows the clear appearance of
an insulating gap.}
\end{figure}

\section{Onsite Matrices}

The on-site term ($H_{\rm CF}$) is written in second quantized language as
$$ H_{\rm CF} = \sum_{\ell \ell'} A_{\ell m} \sum_{s} c^\dagger_{\ell s}c_{m s} $$
where $\ell$ and $m$ are orbital indices, and $s$ refers to the electron spin. Below, we present the $A$ matrices for each of the materials considered in the main text. All the matrices below are written in the orbital basis $\{ (yz , xz , xy) , (x^2\!-\!y^2 , 3z^2\!-\!1) \}$

\subsection{CoTiO$_3$}

\bea
A_{\rm CTO} = \begin{pmatrix}
 -1.1642 &&  0.0312 && -0.0008 && -0.0063 && -0.0039 \\
  0.0312 && -0.1429 && -0.0040  &&  0.0033 && -0.0044 \\
 -0.0008 && -0.004  && -1.0968 &&  0.0926 &&  0.0103 \\
 -0.0063 &&  0.0033 &&  0.0926 && -0.4133 &&  0.4380 \\
 -0.0039 && -0.0044 &&  0.0103 &&  0.4380  && -0.8765
\end{pmatrix}
\eea

\subsection{BaCo$_2$(AsO$_4$)$_2$}

\bea
A_{\rm BCAO} = \begin{pmatrix}
 -2.9578 &&  0.0140 &&  -0.0005 &&  0.0202 &&  0.0123 \\
  0.0140 &&  -1.6838 &&  0.0124 &&  0.0020 &&  -0.0043 \\
 -0.0005 &&  0.0124 && -2.8610 &&   0.049 &&  -0.0495 \\
  0.0202 &&  0.0020 &&   0.0490 &&  -1.9904 && 0.5312 \\
  0.0123 && -0.0043 && -0.04950 && 0.5312 && -2.6055
\end{pmatrix}
\eea

\subsection{BaCo$_2$(PO$_4$)$_2$}

\bea
A_{\rm BCPO} = \begin{pmatrix}
 -3.5049 && -0.0717  && 0.0008 && 0.0034 && 0.0008 \\
 -0.0717 && -2.2912  && 0.0017 && -0.0136 &&  0.0249 \\
  0.0008 &&  0.0017 && -3.4155 && -0.0900  &&   0.0157 \\
  0.0034 && -0.0136 && -0.0900 &&  -2.5757 && 0.5051 \\
  0.0008 &&  0.0249 &&  0.0157 &&  0.5051 && -3.186 \\
\end{pmatrix}
\eea

\section{Hopping Matrices}

Below, we list the hopping matrices for the relevant bonds, listing both the connecting vector $\vec{V}$, as well as the associated hopping matrix $T$. These should be read in terms of the hopping Hamiltonian (in the same basis as above)

\bea
\hat{T}_{i , i+\vec{V}} = \sum_{\ell\ell'\alpha}(T_{\ell\ell'} c^\dg_{i+\vec{V} , \ell'\alpha} c^\pdg_{i\ell\alpha} + T^t_{\ell'\ell} c^\dg_{i\ell\alpha} c^\pdg_{i+\vec{V} , \ell'\alpha})
\eea

\subsection{CoTiO$_3$}

\subsubsection{In-plane 1st Nearest Neighbour}

\bea
\vec{V} = (-0.5 , -0.866 , 0.205) \quad , \quad
T = \begin{pmatrix}
 -0.0803 && -0.0411 &&  0.1233 && -0.0163 && -0.0415 \\
 -0.0411 && -0.0588 &&  0.0236 && -0.0143 && -0.0433 \\
  0.1233 &&  0.0236 && -0.0684 &&  0.0575 && -0.0073 \\
 -0.0163 && -0.0143 &&  0.0575 && -0.0333 && -0.0641 \\
 -0.0415 && -0.0433 && -0.0073 && -0.0641 && -0.0093
\end{pmatrix}
\eea

\bea
\vec{V} = (1.0 , 0.0 , 0.205) \quad , \quad
T = \begin{pmatrix}
  0.0012 &&  -0.0100 && -0.0026 && -0.0243 && -0.0134 \\
 -0.0100 &&  -0.0427 &&  0.0194 &&  0.0268 && -0.0026 \\
 -0.0026 &&  0.0194 &&  0.0793 && -0.0515 && -0.0051 \\
 -0.0243 &&  0.0268 && -0.0515 && -0.1734 &&  0.0811 \\
 -0.0134 && -0.0026 && -0.0051 &&  0.0811 && -0.1146
\end{pmatrix}
\eea

\bea
\vec{V} = (-0.5 , 0.866 , 0.205) \quad , \quad
T = \begin{pmatrix}
 -0.0779 && -0.0214 && -0.1259 && -0.0006 &&  0.0287 \\
 -0.0214 && -0.0991 && -0.0691 &&  0.0119 &&  0.0123 \\
 -0.1259 && -0.0691 && -0.0744 &&  0.0405 && -0.0154 \\
 -0.0006 &&  0.0119 &&  0.0405 && -0.0364 && -0.0271 \\
  0.0287 &&  0.0123 && -0.0154 && -0.0271 &&  0.0376
\end{pmatrix}
\eea

\subsubsection{Out-of-plane 1st Nearest Neighbour}

\bea
\vec{V} = (0.0 , 0.0 , 1.381) \quad , \quad
T = \begin{pmatrix}
 -0.0015 &&  0.0210  &&  0.0010 &&   0.0082 &&  0.0052 \\
  0.0210  &&  0.027  &&  0.0052 && -0.0049 &&  0.0067 \\
  0.0010  &&  0.0052 && -0.019 &&  -0.0097 &&  0.0033 \\
  0.0082 && -0.0049 && -0.0097 &&  0.0353 &&  0.0086 \\
  0.0052 &&  0.0067 &&  0.0033 &&  0.0086 && -0.0142
\end{pmatrix}
\eea

\subsubsection{Out-of-plane 2nd Nearest Neighbour}

\bea
\vec{V} = (-0.5 , -0.866 , 1.587) \quad , \quad
T = \begin{pmatrix}
 -0.0089 && -0.0308 && -0.0046 && -0.0155 &&  0.0042 \\
  0.0358 &&  0.0038 &&  0.0106 &&  0.0063 && -0.0129 \\
  0.0131 &&  0.0124 &&  0.0070 &&   0.0079 &&  -0.0065 \\
  0.0208 &&  0.0234 && -0.0049 &&  0.0144 && -0.0140 \\
  0.0005 &&  0.0276 && -0.0167 &&  0.0264 && -0.0216
\end{pmatrix}
\eea

\bea
\vec{V} = (0.5 , 0.866 , -1.587) \quad , \quad
T = \begin{pmatrix}
 -0.0089 &&  0.0358 &&  0.0131 &&  0.0208 &&  0.0005 \\
 -0.0308 &&  0.0038 &&  0.0124 &&  0.0234 &&  0.0276 \\
 -0.0046 &&  0.0106 &&  0.0070 && -0.0049 && -0.0167 \\
 -0.0155 &&  0.0063 &&  0.0079 &&  0.0144 &&   0.0264 \\
  0.0042 && -0.0129 && -0.0065 && -0.0140 &&  -0.0216
\end{pmatrix}
\eea

\bea
\vec{V} = (1.0 , 0.0 , 1.587) \quad , \quad
T = \begin{pmatrix}
  0.0109 && -0.0542 &&  0.0071 && -0.0099 && -0.0169 \\
  0.0275 &&  0.0071 &&  0.0056 && -0.0019 && -0.0022 \\
 -0.0295 &&  0.0188 && -0.0137 &&  0.0062 &&  0.0201 \\
 -0.0036 && -0.0051 && -0.0085 &&  0.0013 &&  0.0184 \\
  0.0104 && -0.0117 &&  0.0016 &&  0.0017 && -0.0109
\end{pmatrix}
\eea

\bea
\vec{V} = (-1.0 , 0.0 , -1.587) \quad , \quad
T = \begin{pmatrix}
  0.0109 &&  0.0275 && -0.0295 && -0.0036 &&  0.0104 \\
 -0.0542 &&  0.0071 &&  0.0188 && -0.0051 && -0.0117 \\
  0.0071 &&  0.0056 && -0.0137 && -0.0085 && 0.0016 \\
 -0.0099 && -0.0019 &&  0.0062 &&  0.0013 &&  0.0017 \\
 -0.0169 && -0.0022 &&  0.0201 &&  0.0184 && -0.0109
\end{pmatrix}
\eea

\bea
\vec{V} = (-0.5 , 0.866 , 1.587) \quad , \quad
T = \begin{pmatrix}
  0.0174 && -0.0180 && 0.0049 && -0.0224 &&  0.0102 \\
  0.0232 && -0.0098 && 0.0085 && -0.0022 &&  0.0031 \\
 -0.0146 && -0.0278 && 0.0052 &&  0.0079 && -0.0013 \\
  0.0534 &&  0.0160 &&  0.0090 &&  -0.0153 &&  0.0179 \\
  0.0093 && -0.0087 && 0.0063 &&  0.0024 && -0.0027
\end{pmatrix}
\eea

\bea
\vec{V} = (0.5 , -0.866 , -1.587) \quad , \quad
T = \begin{pmatrix}
  0.0174 &&  0.0232 && -0.0146 &&  0.0534 &&  0.0093 \\
 -0.0180 && -0.0098 && -0.0278 &&  0.0160 &&  -0.0087 \\
  0.0049 &&  0.0085 &&  0.0052 &&  0.0090 &&   0.0063 \\
 -0.0224 && -0.0022 &&  0.0079 && -0.0153 &&  0.0024 \\
  0.0102 &&  0.0031 && -0.0013 &&  0.0179 && -0.0027
\end{pmatrix}
\eea

\subsection{BaCo$_2$(AsO$_4$)$_2$}

\subsubsection{In-plane 1st Nearest Neighbour}

\bea
\vec{V} = (-0.5 , -0.866 , 0.056) \quad , \quad
T = \begin{pmatrix}
 -0.1539 && -0.0133 &&  0.1892 && -0.0808 && 0.0095 \\
 -0.0133 && -0.0250  &&  0.0676 && -0.0087 &&  0.0115 \\
  0.1892 &&  0.0676 && -0.0883 &&  0.0240 &&  -0.0012 \\
 -0.0808 && -0.0087 &&  0.0240  &&  0.0292 && -0.056 \\
  0.0095 &&  0.0115 && -0.0012 && -0.056 &&  0.0176 \\
\end{pmatrix}
\eea

\bea
\vec{V} = (1.0 , 0.0 , 0.056) \quad , \quad
T = \begin{pmatrix}
  0.0961 &&  0.0558 &&  0.0009 &&  0.0123 && -0.0033 \\
  0.0558 && -0.0240 &&  -0.0076 &&  0.0129 && -0.0221 \\
  0.0009 && -0.0076 &&  0.0442 &&  0.0052 && -0.0476 \\
  0.0123 &&  0.0129 &&  0.0052 && -0.1668 &&  0.1650 \\
 -0.0033 && -0.0221 && -0.0476 &&  0.1650 &&  -0.1699
\end{pmatrix}
\eea

\bea
\vec{V} = (-0.5 , 0.866 , 0.056) \quad , \quad
T = \begin{pmatrix}
 -0.1581 && -0.0391 && -0.1913 &&  0.0531 && -0.0147 \\
 -0.0391 && -0.0226 && -0.0711 &&  0.0043 && -0.0034 \\
 -0.1913 && -0.0711 && -0.0849 &&  0.0005 && -0.0034 \\
  0.0531 &&  0.0043 &&  0.0005 &&  0.0416 && -0.0649 \\
 -0.0147 && -0.0034 &&  -0.0034 && -0.0649 &&  0.0038
\end{pmatrix}
\eea

\subsubsection{In-plane 3rd Nearest Neighbour}

\bea
\vec{V} = (1.0 , 1.732 , 0.056) \quad , \quad
T = \begin{pmatrix}
 -0.0177 && -0.0075 &&  0.0255 && -0.0102 && -0.0026 \\
 -0.0075 &&  0.0013 &&  0.0298 && -0.0658 && -0.0377 \\
  0.0255 &&  0.0298 && -0.0283 && -0.0060  && -0.0005 \\
 -0.0102 && -0.0658 && -0.0060  &&  0.0653 &&  0.0394 \\
 -0.0026 && -0.0377 && -0.0005 &&  0.0394 &&  0.0251
\end{pmatrix}
\eea

\bea
\vec{V} = (-2.0 , 0.0 , 0.056) \quad , \quad
T = \begin{pmatrix}
  0.0061 &&  0.0137 && -0.0007 &&  0.0077 &&  0.0000 \\
  0.0137 &&  0.1327 && -0.0134 && -0.0042 &&  0.0078 \\
 -0.0007 && -0.0134 && -0.0027 && -0.0086 &&  0.0015 \\
  0.0077 && -0.0042 && -0.0086 && -0.0646 &&  0.0149 \\
  0.0000 && 0.0078 &&  0.0015 &&  0.0149 && -0.0258
\end{pmatrix}
\eea

\bea
\vec{V} = (1.0 , -1.732 , 0.056) \quad , \quad
T = \begin{pmatrix}
 -0.0226 &&  -0.0107 && -0.0243 && 0.0141 && 0.0094 \\
 -0.0107 &&  0.0019 && -0.0084 && 0.0661 && 0.0363 \\
 -0.0243 && -0.0084 && -0.0242 && 0.0129 && 0.0151 \\
  0.0141 &&  0.0661 &&  0.0129 && 0.0856 && 0.0260 \\
  0.0094 &&  0.0363 &&  0.0151 && 0.0260 &&  0.0051
\end{pmatrix}
\eea

\subsection{BaCo$_2$(PO$_4$)$_2$}

\subsubsection{In-plane 1st Nearest Neighbour}

\bea
\vec{V} = (-0.5 , -0.866 , 0.059) \quad , \quad
T = \begin{pmatrix}
  0.1162 && -0.0458 && 0.0024 && -0.0152 &&  0.0031 \\
 -0.0458 && -0.0748 && 0.0106 &&  0.0078 && -0.0217 \\
  0.0024 &&  0.0106 && 0.0363 &&  0.0057 &&  0.0519 \\
 -0.0152 &&  0.0078 && 0.0057 && -0.1642 &&  0.1553 \\
  0.0031 && -0.0216 && 0.0519 &&  0.1553 &&  -0.2813
\end{pmatrix}
\eea

\bea
\vec{V} = (1.0 , 0.0 , 0.059) \quad , \quad
T = \begin{pmatrix}
 -0.2091 &&  0.0040 &&  -0.2255 && -0.0496 &&  0.0435 \\
  0.0040  && -0.0612 &&  0.0309 &&  0.0033 &&  -0.0031 \\
 -0.2255 &&  0.0309 && -0.0987 &&  0.0137 &&  0.0194 \\
 -0.0496 &&  0.0033 &&  0.0137 &&  0.0115 && -0.0874 \\
  0.0435 && -0.0031 &&  0.0194 && -0.0874 && -0.0103
\end{pmatrix}
\eea

\bea
\vec{V} = (-0.5 , 0.866 , 0.059) \quad , \quad
T = \begin{pmatrix}
 -0.1827 && -0.0309 &&  0.2215 &&  0.0908 && -0.0315 \\
 -0.0309 && -0.0711 && -0.0265 &&  0.0043 &&  0.0059 \\
  0.2215 && -0.0265 &&  -0.1217 && -0.0213 &&  0.0096 \\
  0.0908 &&  0.0043 && -0.0213 && -0.0084 && -0.0761 \\
 -0.0315 &&  0.0059 &&  0.0096 && -0.0761 &&  0.0161
\end{pmatrix}
\eea

\subsubsection{In-plane 3rd Nearest Neighbour}

\bea
\vec{V} = (1.0 , 1.732 , 0.056) \quad , \quad
T = \begin{pmatrix}
  0.0243 && -0.0360 &&  -0.0014 && -0.0084 && -0.0044 \\
 -0.0360 &&  0.1278 &&  0.0189 && -0.0055 &&  0.0141 \\
 -0.0014 && 0.0189 && -0.0197 &&  0.0236 && -0.0042 \\
 -0.0084 && -0.0055 &&  0.0236 && -0.0463 &&  0.0087 \\
 -0.0044 && 0.0141 && -0.0042 &&  0.0087 && -0.0275
\end{pmatrix}
\eea

\bea
\vec{V} = (-2.0 , 0.0 , 0.056) \quad , \quad
T = \begin{pmatrix}
 -0.0118 && -0.0024 && -0.0208 && -0.0211 && -0.0054 \\
 -0.0024 &&  0.0038 &&  0.0056 &&  0.0639 &&  0.0348 \\
 -0.0208 &&  0.0056 && -0.0276 && -0.0066 && -0.0233 \\
 -0.0211 &&  0.0639 && -0.0066 &&  0.1013 &&  0.0172 \\
 -0.0054 &&  0.0348 && -0.0233 &&  0.0172 && -0.0072
\end{pmatrix}
\eea

\bea
\vec{V} = (1.0 , -1.732 , 0.056) \quad , \quad
T = \begin{pmatrix}
 -0.0062 && -0.0117 &&  0.0222 &&  0.0200 &&  0.0014 \\
 -0.0117 &&  0.0059 && -0.0292 && -0.0659 && -0.0358 \\
  0.0222 && -0.0292 && -0.0342 &&  0.0151 && -0.0116 \\
  0.0200 && -0.0659 &&  0.0151 &&  0.0809 &&  0.0298 \\
  0.0014 && -0.0358 && -0.0116 &&  0.0298 &&  0.0121
\end{pmatrix}
\eea

\section{Dependence of Exchanges on Interaction Parameters}

In this section, we give more examples of the dependence of the exchanges considered in the main text on the interaction parameters. This is done to emphasize the fact that for reasonable values of these parameters, there is a fairly large variance in the exchanges.

\subsection{CoTiO$_3$}
\begin{figure}[H]
  \centering
  \includegraphics[width=.5\textwidth]{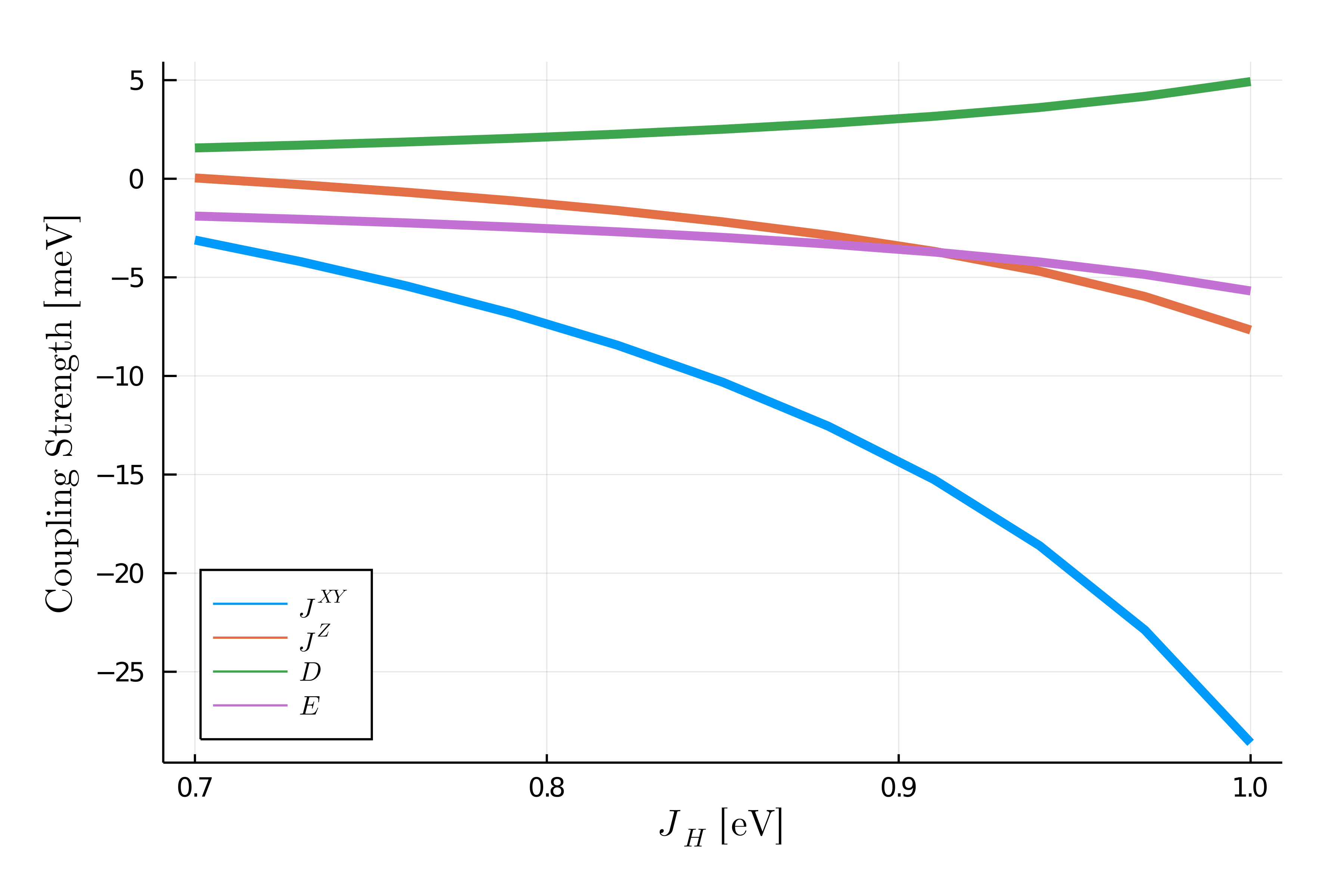}
  \caption{Variance of 1NN Exchanges with Hund's Coupling $J_H$}
\end{figure}

\subsection{BaCo$_2$(AsO$_4$)$_2$ and BaCo$_2$(PO$_4$)$_2$}

The Barium compounds show similar behaviour, so the results will only be presented for BaCo$_2$(AsO$_4$)$_2$.

\begin{figure}[h]
  \centering
  \includegraphics[width=.5\textwidth]{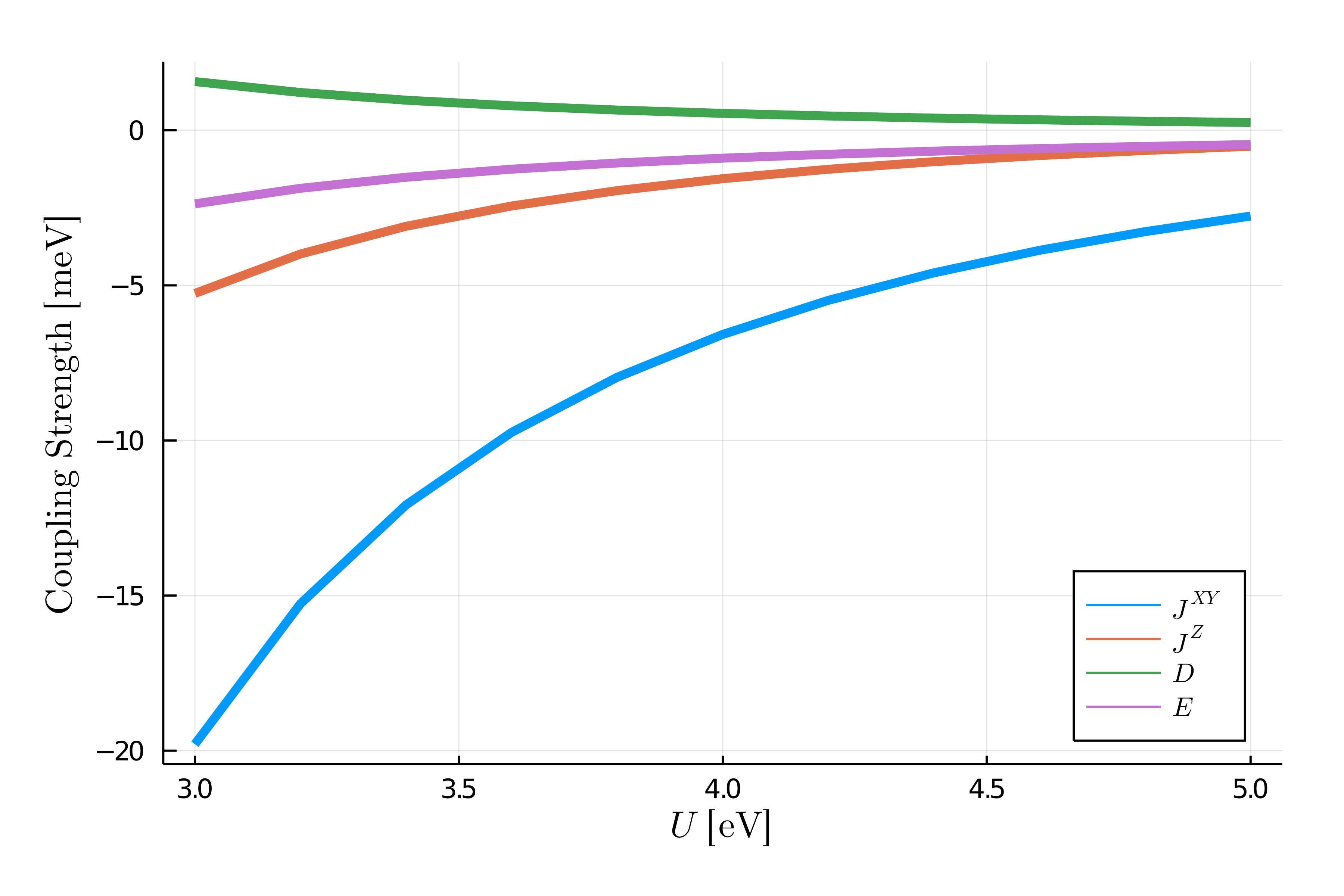}
  \caption{Variance of 1NN Exchanges with Hubbard $U$}
\end{figure}
\begin{figure}[h]
  \centering
  \includegraphics[width=.5\textwidth]{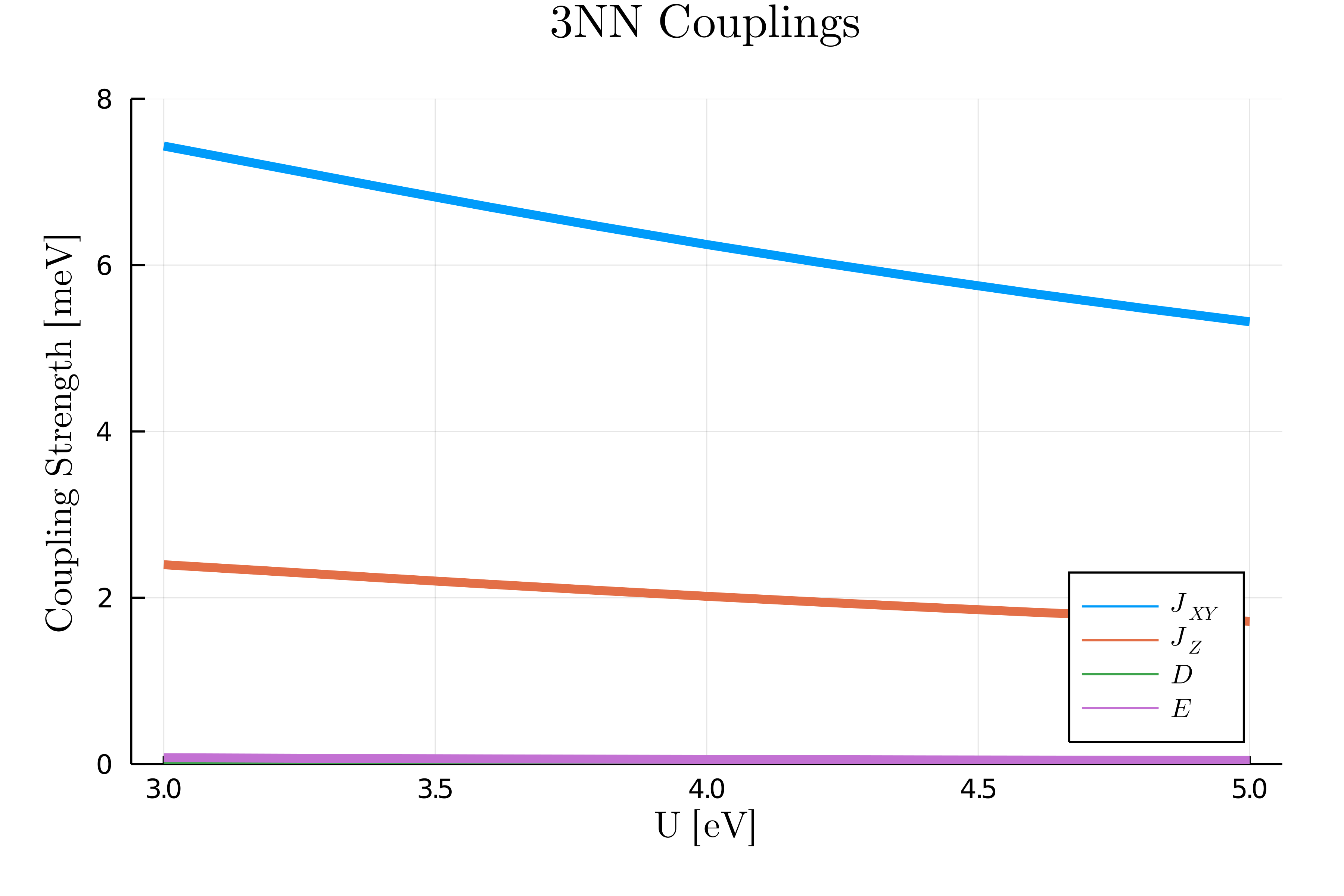}
  \caption{Variance of 3NN Exchanges with Hubbard $U$}
\end{figure}
\begin{figure}[h]
  \centering
  \includegraphics[width=.5\textwidth]{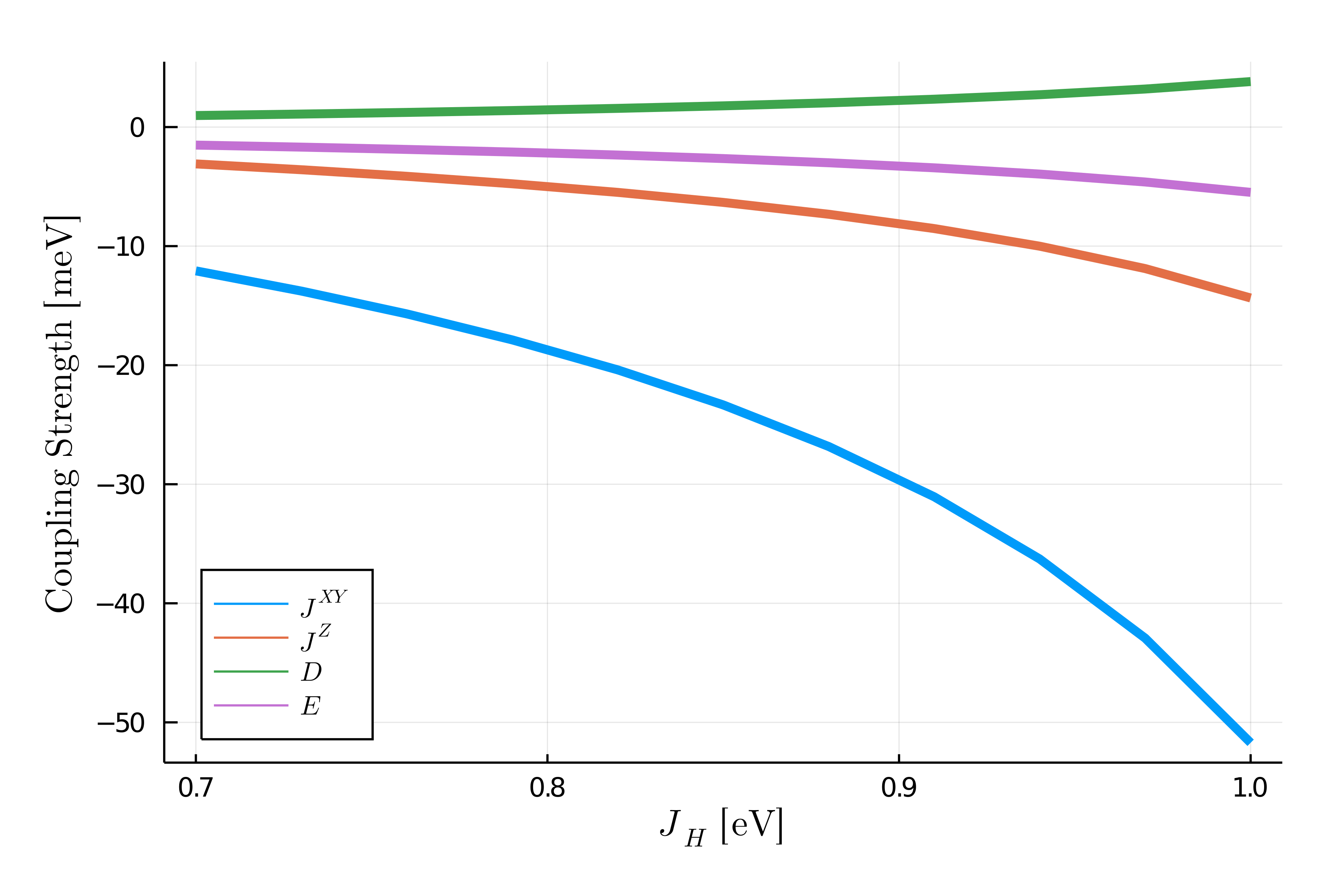}
  \caption{Variance of 1NN Exchanges with Hund's coupling $J_H$}
\end{figure}
\begin{figure}[h]
  \centering
  \includegraphics[width=.5\textwidth]{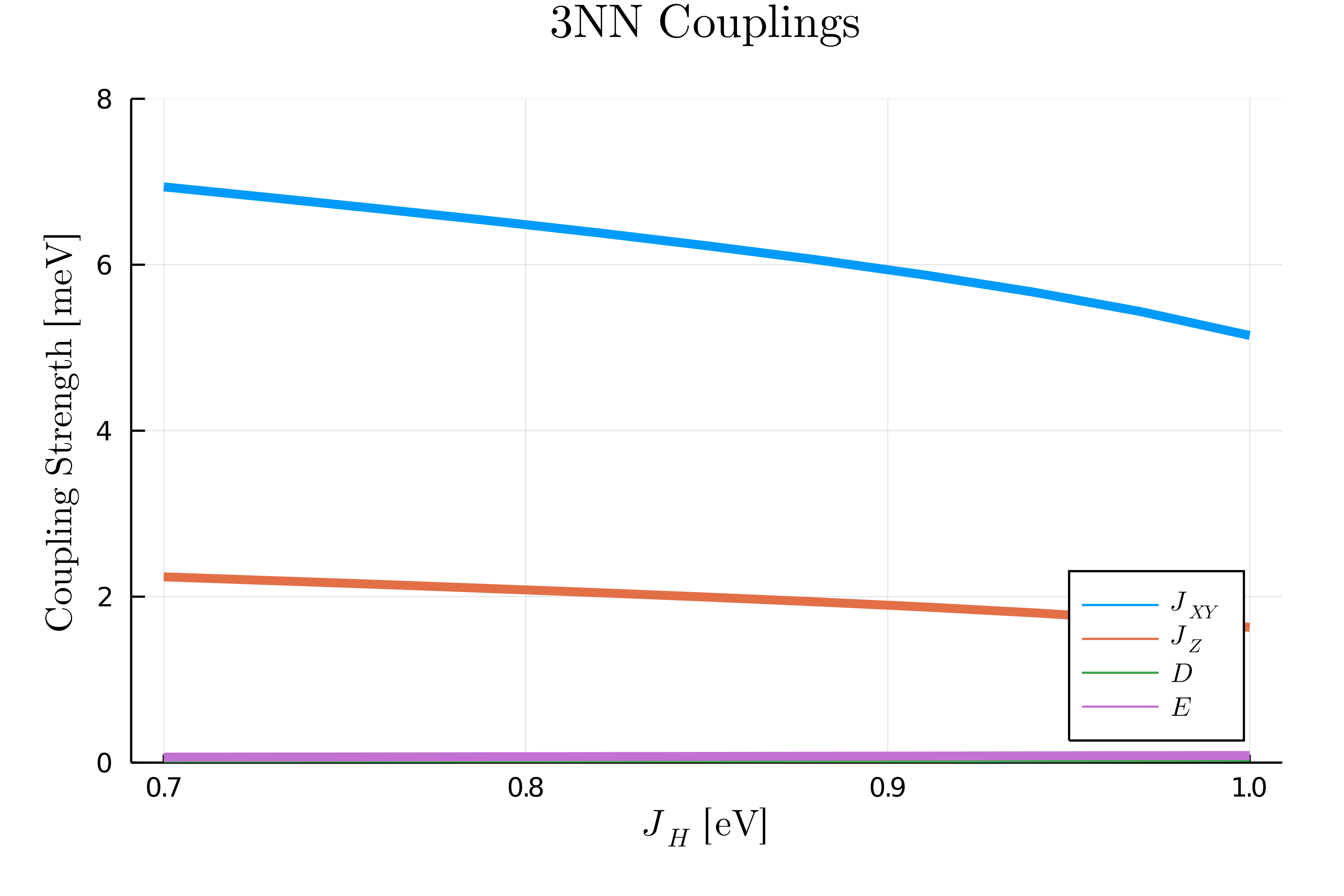}
  \caption{Variance of 3NN Exchanges with Hund's coupling $J_H$}
\end{figure}
\begin{figure}[h]
  \centering
  \includegraphics[width=.5\textwidth]{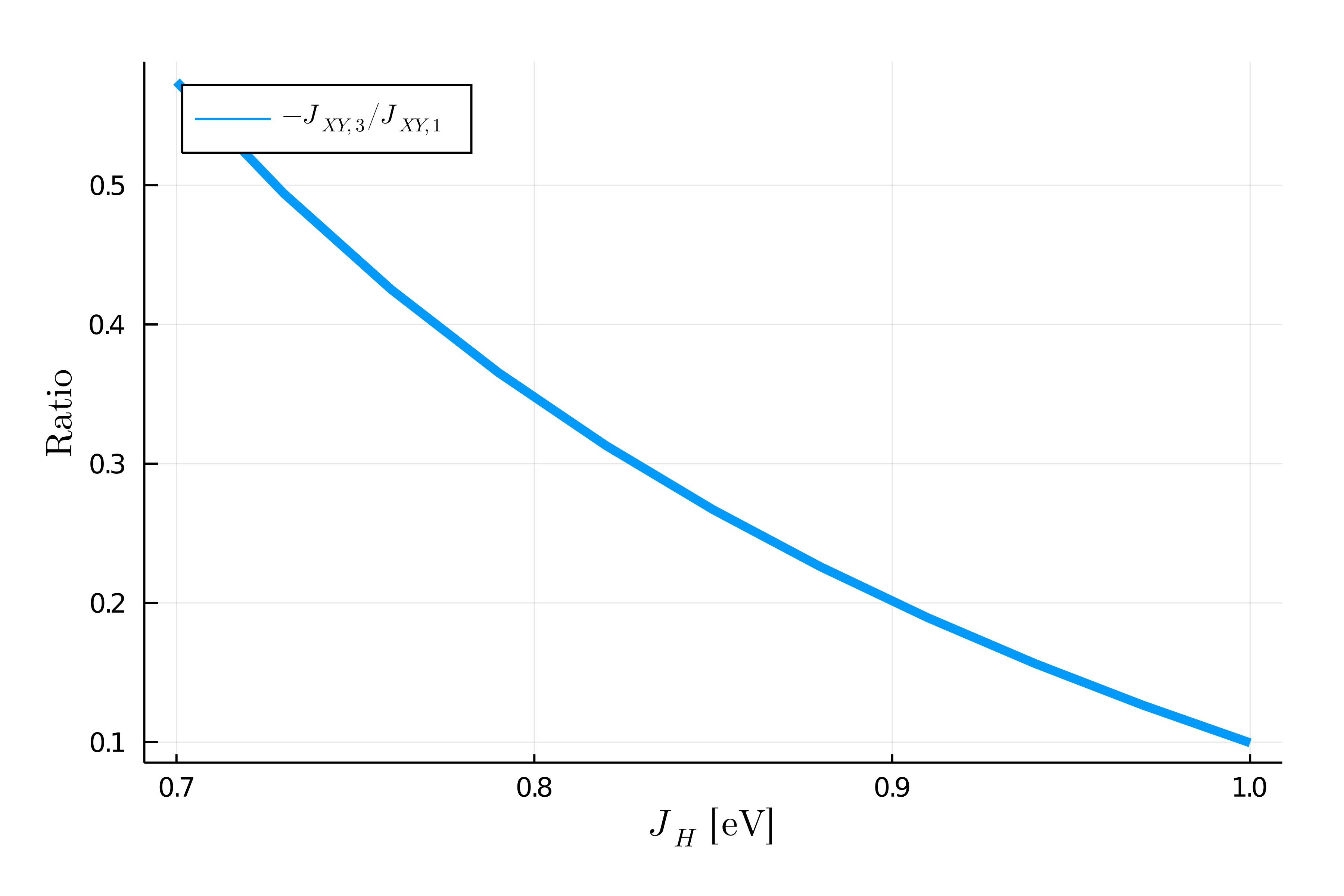}
  \caption{Variance of the ratio $-J_3^{XY}/J_1^{XY}$ with Hund's coupling $J_H$}
\end{figure}

\clearpage

\bibliography{d7exchange}

%
%
%
%
%
%
%